\documentclass{aa} 

\usepackage{graphicx}
\usepackage{natbib}
\usepackage{txfonts}
\usepackage[draft,breaklinks]{hyperref}
\newcommand{\mygi}{MyGIsFOS}
\newcommand{\teff}{\ensuremath{T_\mathrm{eff}}}

\newcommand{\kms}{$\rm km\,s ^{-1}$}
\bibpunct{(}{)}{;}{a}{}{,}
\begin{document} 

\title{ TOPoS VI. The metal-weak tail of the  metallicity distribution functions of
the Milky Way and the Gaia-Sausage-Enceladus structure \thanks{Spectroscopic and photometric metallicities
derived and discussed in this paper 
as well as orbital actions computed and discussed this paper 
are only available in electronic form
at the CDS via anonymous ftp to cdsarc.u-strasbg.fr (130.79.128.5)
or via \url{http://cdsweb.u-strasbg.fr/cgi-bin/qcat?J/A+A/}}
}
\titlerunning{MDF}
\author{
P. Bonifacio\inst{1} \and
L. Monaco\inst{2} \and
S. Salvadori\inst{3,4} \and
E. Caffau\inst{1} \and
M. Spite\inst{1} \and
L. Sbordone \inst{5} \and
F. Spite \inst{1} \and
H.-G.~Ludwig \inst{6} \and
P.~Di~Matteo \inst{1} \and
M.~Haywood \inst{1} \and
P.~Fran{\c c}ois \inst{1,7} \and
A.J.~Koch-Hansen\inst{8}\and
N.~Christlieb\inst{6}
\and
S.~Zaggia\inst{9}
}
\institute{
GEPI, Observatoire de Paris, Universit\'e PSL, CNRS, Place Jules
Janssen, F-92195 Meudon, France\and
Departamento de Ciencias Fisicas, Universidad Andres Bello, Fernandez Concha 700
, Las Condes, Santiago, Chile\and
Dipartimento di Fisica e Astronomia, Universit\'a degli Studi di Firenze, 
Via G. Sansone 1, I-50019 Sesto Fiorentino, Italy\and
INAF/Osservatorio Astrofisico di Arcetri, Largo E. Fermi 5, I-50125
Firenze, Italy \and
European Southern Observatory, Casilla 19001, Santiago, Chile\and
Zentrum f\"ur Astronomie der Universit\"at Heidelberg, Landessternwarte, K\"onigstuhl 12, 69117 Heidelberg, Germany \and
UPJV, Universit\'e de Picardie Jules Verne, 33 rue St Leu, 80080 Amiens, France \and
Zentrum f\"ur Astronomie der Universit\"at Heidelberg, Astronomisches Rechen-Institut, M\"onchhofstr. 12, 69120 Heidelberg, Germany
\and 
INAF—Osservatorio Astronomico di Padova, Vicolo dell'Osservatorio 5, Padova, I-35122, Italy
}

 \date{Received ; accepted }
\abstract
{ The goal of the Turn-Off Primordial Stars survey  (TOPoS) 
project is to find and analyse turn-off (TO) stars of extremely low
metallicity. To select the targets for spectroscopic follow-up at high spectral
resolution, we relied on low-resolution spectra from the Sloan Digital Sky Survey (SDSS).}
{In this paper, we use the metallicity estimates  
we obtained from our analysis
of the SDSS spectra to construct the metallicity distribution function (MDF) of the Milky
Way, with special emphasis on its metal-weak tail. The goal
is to provide the underlying distribution out of which the TOPoS sample was extracted.}
{We made use of SDSS photometry, Gaia photometry, and distance estimates derived from the Gaia
parallaxes to derive a metallicity estimate for a large sample of over 24 million TO stars. 
This sample was used to derive the metallicity bias of the sample for which SDSS spectra are available.}
{We determined that the spectroscopic sample is strongly biased in favour of metal-poor stars, as intended.
A comparison with the unbiased photometric sample allows us to correct for the selection bias. 
We selected a sub-sample of stars with reliable parallaxes for which we combined the SDSS radial velocities
with Gaia proper motions and parallaxes to compute actions and orbital
parameters in the Galactic potential. This allowed us to characterise
the stars dynamically, and in particular to select a sub-sample that belongs to the
Gaia-Sausage-Enceladus (GSE) accretion event. We are thus also able to provide the MDF of  GSE.}
{The metal-weak tail derived in our study is very similar to that derived in the H3 survey and
in the Hamburg/ESO Survey. This allows us to average the three MDFs and provide an error bar
for each metallicity bin. Inasmuch  as the GSE structure is representative
of the progenitor galaxy that collided with the Milky Way, that
galaxy appears to be strongly deficient in metal-poor stars compared
to the Milky Way, suggesting that the metal-weak tail of the latter has been largely formed
by accretion of low-mass galaxies rather than massive galaxies, such as the GSE progenitor.}
   \maketitle

\section{Introduction}

Considerable information on a galaxy's formation and evolution is
imprinted in its metallicity distribution function (MDF).
To our knowledge, the first attempt to determine the MDF of the Galaxy
was that of \citet{vdb62}.  Since \citet{Wallerstein62} robustly
established the correlation between metallicity and ultraviolet
excess, $\delta(U-B)$, \citet{vdb62} proceeded to study the
distribution of the ultraviolet excess of a sample of 142 stars, finding
that metal-poor stars are very rare, and that no star in his sample had
a metallicity below $-0.85$\footnote{One should be aware that in the
  literature `metallicity' may refer to slightly different
  quantities; often it is [Fe/H], especially in the older papers, but
  not always. In Sect.\,\ref{sec:sel}, we provide the definition
  of metallicity used in this paper.}.
It was quickly appreciated that the ultraviolet excess, as a metallicity
indicator, saturates around a metallicity of $-2.0$, thus calling for
spectroscopic metallicities, or photometric indices based on
intermediate or narrow-band photometry.  Simple models of Galactic
chemical evolution and the long life time of low-mass stars led to
the expectation that it should be possible to identify stars below $-2.0$
and $-3.0$ in dedicated observational programmes, and hence considerable
effort was devoted specifically to the discovery of metal-poor stars.

Extensive objective-prism surveys \citep{Bond70,Bidelman73,Bond80}
failed to detect stars below --3.0 and led \citet{Bond81} to
conclude that `very few stars with [Fe/H]$< -3$ may exist. Clearly one
  must go to quite faint objects before encountering such Population
  III stars'\footnote{\citet{Bond81} refers to any star with
    [Fe/H]$<-3$ as Population III.}, in spite of the fact that
\citet{Bessel77} had already claimed $\mathrm{[Ca/H]}= -3.5$ for the
giant CD~$-38^\circ \, 245$, based on a low-resolution spectrum.

\citet{Hartwick76} made use of spectroscopic metallicities determined
for a few individual stars in several globular clusters and showed
that their MDF extends at least down to a metallicity of $-2.2$.
\citet{Cayrel86} showed that if star formation started on a gas cloud
of $10^6 M_{\sun}$, the O-type stars that formed would rapidly evolve and
enrich the cloud to a level of $Z\simeq 0.2\,Z_{\sun}$, thus explaining
the paucity of metal-poor stars. This scenario is still considered
valid today, modulo the introduction of a dark matter halo, which
allows us to relax the hypothesis on the baryonic mass.  The question
remains, however, as to whether low-mass
stars were also formed at the same time as O-stars.

\citet{Beers85} demonstrated the efficiency of coupling a wide-field
objective-prism survey with low-resolution ($R\sim
2\ 000$)\ spectroscopic follow-up. From the low-resolution spectra,
they were able to extract metallicities using a spectral index centred on
the \ion{Ca}{ii} H\&{K} lines, which resulted in a catalogue of 134
metal-poor stars.  This sample extended down to metallicity $-4.0$
(star CS\,22876-32). The MDF was not explicitly shown in that paper,
but it was concluded that the slope of the metal-poor tail of the MDF
is approximately constant. \citet{Beers87} presented the MDF of that
sample of 134 stars. The success of \citet{Beers85} in finding
extremely metal-poor stars on one hand lies in the fainter
magnitude limit of their survey ($B \sim 15$, compared to $B \sim
11.5$ of \citealt{Bond80}), and on the other hand in the choice of
shortening the spectra on the plate by using an interference filter
that selected only the UV-blue wavelengths, reducing the fraction of
overlapping spectra, and also minimising the diffuse sky background,
hence allowing for a fainter magnitude limit.

\citet{Carney87} demonstrated how metallicities could be reliably
derived from high-resolution (R$\sim 30\ 000$), low signal-to-noise
spectra, resulting in a catalogue of 818 stars that provided an MDF
extending to $-3.0$ \citep{Laird88}.  Presumably that was the first
time at which the derived MDF clearly displayed a bi-modality,
although \citet{Laird88} did not comment on it.

\citet{Ryan91a} used low-resolution spectroscopy and defined a
spectral index, centred on \ion{Ca}{ii} H\& K lines, slightly
different from that defined by \citet{Beers85}, and they obtained
metallicities for a sample of 372 kinematically selected halo
stars. The MDF was presented and discussed in \citet{Ryan91}. The MDF
extends down to $\sim -4.0$ and is not bi-modal, this can however
be attributed to the kinematical selection that excluded thick disc
stars.

In the last twenty years, there have been many attempts to define the
MDF of the Galactic halo, a non-exhaustive list includes
\citet{Schoerck}, { \citet{Yong13}}, \citet{Allende14},\citet{Youakim}, \citet{Naidu20}, and
\citet{Carollo20}, and we compare our results with some of theirs.
Our group has long been exploiting the spectra of the Sloan Digital Sky
Survey (hereafter SDSS, \citealt{sdss,DR6,DR9,DR12}) to select
extremely metal-poor stars
\citep{xgto,bonifacio11,leo,bonifacio12,xgto13,topos1,toposII,toposIII,toposIV,francois18,toposV,
  francois2020}.  
The TOPoS project \citep{topos1} was designed to exploit the
high success rate of the selection of extremely metal-poor stars
to study their detailed abundances.
The  high-resolution follow-up observations
were carried out with XShooter and UVES at the ESO VLT telescope
\citep{toposII,toposIII,toposIV,toposV}.  
The selection technique is described
in \citet{Ludwig08} and \citet{xgto,topos1}. It made use of spectra of
turn off (TO) stars from SDSS~DR6 \citep{DR6}, DR9 \citep{DR9}, and
DR12 \citep{DR12}, selected as described in Appendix \ref{appa}. A
version of the raw MDF from SDSS~DR6 has been shown in
\citet{Ludwig08}, Figure~3. Some interesting features were already
apparent, including the bi-modality of the MDF, which is similar to that of
\citet{Laird88} identified at the time as 'halo' and `thick
disc'.  However, it was clear that the metal-weak tail of the raw MDF
was unrealistic, providing a flat MDF at low metallicity.  Visual
inspection of the spectra allowed us to identify many white dwarfs
with a \ion{Ca}{ii}~K line absorption (likely due to circumstellar
material or material accreted from a planet orbiting the white
dwarf). However, we were unable at that time to robustly discriminate
against these objects by an automated analysis of the spectra, even
when also using photometric data. In addition, we were well aware that
the selection function of the SDSS spectra we were using was difficult
to define.  For these reasons, we have previously not published an analysis
of the MDF derived from our analysis of SDSS spectra.

The situation changed dramatically with the Gaia \citep{Gaia} Data
Release 2 \citep{GaiaDR2,Arenou18}. The Gaia photometry and astrometry
allowed us to remove the white dwarfs contaminating our sample, and to
clean the sample from stars with inaccurate SDSS photometry.
Distances based on Gaia parallaxes allowed us to also derive a
photometric metallicity estimate from SDSS photometry. This was used
to quantify the metallicity bias present in our SDSS spectroscopic
sample.
With these improvements, we now present our estimate of the MDF. 
In Sect.\,\ref{sec:mdf}, we also provide a 
detailed comparison with some of the previously published MDFs.

\section{Selection and analysis of SDSS spectra\label{sec:sel}}

We selected from SDSS data release 12 \citep{DR12} stars with TO
colours, obtaining a set of 312\,009 spectra of 266\,442 unique stars
(see Appendix \ref{appa}).  
{ The average S/N of the spectra at the effective wavelength of
the $g$ band is $24\pm 17$. 
Both the SDSS and BOSS spectrographs have a red
and a blue arm. The resolving power, $\lambda/\Delta\lambda$,
is 1850\ in the blue and 2200\ in the red for the SDSS spectrograph,
and 1560\ in the blue and 2300\ in the red for the BOSS spectrograph.}
We used the reddening provided in the SDSS
catalogue, which is derived from the maps of \citet{Schlegel}. { At
  the time the data were retrieved from the SDSS archive, as detailed
  in Appendix \ref{appa}, Gaia distances were not yet available, nor
  were 3D maps based on these distances. Our sample spans a distance range
  of roughly 1 to 5\, kpc. Available 3D maps
  \citep[e.g.][]{Lallement} are limited to 3\,kpc, and in many
  directions at high galactic latitude, only to less than
  1\,kpc. Usage of those maps would therefore introduce a distance-
  and coordinate-dependent bias to the sample.}  All the
aforementioned spectra were analysed with \mygi\ \citep{mygisfos},
using a fixed \teff\, derived from $(g-z)_0$ with a third-order
polynomial fit to the theoretical colour for metallicity $-4.0$, and
an $\alpha$-enhancement of $+0.4$.\footnote{\label{tformula} ${\rm T_{eff}} =
  7.2165\times10^3-3.2366\times10^3(g-z)_0+1.1578\times
  10^3(g-z)_0^2-1.8493\times 10^2(g-z)_0^3$}

The surface gravity was kept fixed at $\log g = 4.0$, the
microturbulence was set to 1.5\,km\,s$^{-1}$. The first guess for
each spectrum was a metallicity of $-1.5$. The $\alpha$-element
enhancement was kept fixed at $[\alpha\mathrm{/Fe}]=+0.4$.

\mygi\ used a special grid of synthetic spectra.  The metal-poor part
was computed explicitly using the OSMARCS \citep{G2008} TOPoS grid
\citep[see][for details]{topos1}. This was complemented with 
MARCS interpolated models at higher metallicity. The synthetic spectra were
computed with version 12.1.1 of {\tt turbospectrum}
\citep{alvarez_plez,2012ascl.soft05004P}.

The synthetic spectra contained two wavelength intervals:
380--547\,nm and 845--869\,nm.  They were broadened by 162\,{\kms}
and 136\,{\kms} for the red and blue SDSS spectra, and 192\,{\kms} and
132\,{\kms} for the red and blue BOSS spectra, respectively.
This corresponds to the spectral resolution of the spectra.

A special line list and continuum files were used, in which many
features, including the \ion{Ca}{ii} K and \ion{Ca}{ii} IR triplet
were labelled as {\tt 26.00}, thus treating them as if they were iron.
We deliberately exclude the features due to
C-bearing molecules from our estimate (G-band and Swan band), although the C abundance
is estimated from them.  This was done because the resolution and S/N of the SDSS spectra are not sufficient
for a multi-element analysis, especially at low
metallicity; in many spectra, the only measurable
feature is the \ion{Ca}{ii} K line.  So what \mygi\ provides in this
case is an overall `metallicity' that is essentially a mean of Mg,
Ca, Ti, Mn, and Fe.  { The choice of $\rm [\alpha/Fe]$ only impacts the
  lower metallicities, where the metallicity estimate is dominated by
  the \ion{Ca}{ii} line. At higher metallicities, the Fe feature holds
  more weight, and our estimate is not biased. We verified this by
  looking at our estimates based on the SDSS spectra for a sample of
  seven stars from \citet{caffau18} with low $\rm [\alpha/Fe]$. At low
  metallicity, there is a large scatter, but at metallicity $> -1.0$
  there is a tight correlation with no offset.} Throughout this paper, 
we refer to this quantity as {\em \emph{metallicity}}, and it should
not be confused with $Z$, the mass fraction of metals. Our
metallicity estimate is closer to [Fe/H] than to $Z$, and the
difference can be very large in the case of CEMP stars, which, as
discussed below, we cannot detect in a complete and reliable manner.

It should be kept in mind that our procedure makes a specific assumption of
[$\alpha$/Fe] as a function of [Fe/H]. We assumed $[\alpha\mathrm{/Fe}] =
+0.4$ for $\mathrm{[Fe/H]}\le -0.5$, and $[\alpha\mathrm{/Fe}]=0.0$ otherwise.  Stars
that do not follow this simplistic law will be assigned a wrong
metallicity.  We stress that our analysis was specifically designed to
detect low-metallicity stars for follow-up at higher resolution, and
in this respect it has been extremely successful. In
Sect.\,\ref{sec:pa}, we show that it nevertheless provides a useful metallicity
estimate at higher metallicities as well.

The combination of the low resolution and low S/N of the SDSS spectra
implies that C can be measured reliably only when its spectral features are
very strong. We did not make any attempt to remove CEMP stars
from our sample, since this would result in a strong bias.  Many CEMP
stars would be missed simply because the S/N of the spectrum is
too low. The drawback is that for some (but not all) CEMP stars our
metallicity estimate is far too high, due to the presence of strong
lines of C-bearing molecules, as we show in Sect.\,\ref{sec:pa}

\begin{figure}
\centering
\resizebox{7.5cm}{!}{
\includegraphics{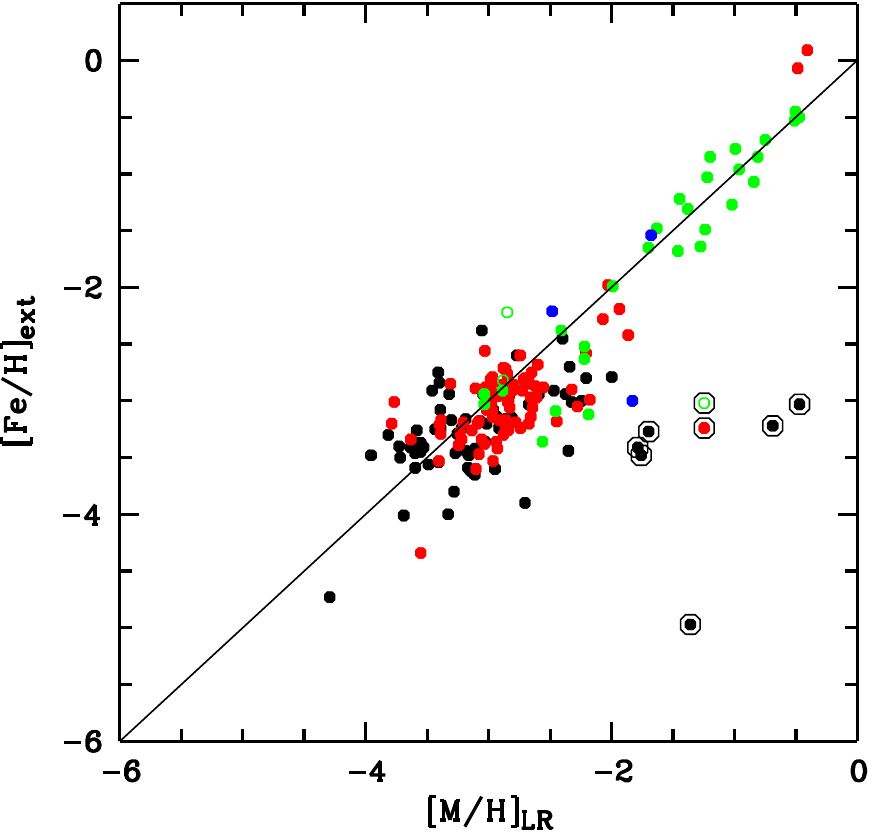}}
\caption{Metallicities derived from SDSS spectra (LR) compared to
  those of independent analysis of the same stars (ext).  The black
  dots are the TOPoS sample (see text); the data are provided in Table
  \ref{hr}. The red dots are the PASTEL sample (see text); the data is
  provided in Table \ref{hrp}. The green dots are data from
  \citet{caffau18}, and open symbols are stars with uncertain abundance
  according to \citet{caffau18}. The blue dots are data from
  \citet{caffau20}. The stars circled in black are removed from the
  average (see text).}
\label{pl_acc}
\end{figure}

\section{Precision and accuracy of the \mygi\ metallicities\label{sec:pa}}

\subsection{Precision \label{metprecision}}

We took advantage of the fact that  15\,496 stars of our
sample have multiple spectra (i.e. up to 33) to estimate the
precision of our metallicities.  The root-mean-square error of the
derived metallicity for stars with multiple spectra is as large as
1.3\,dex; however, its mean value is 0.10\,dex, with a standard
deviation of 0.11\,dex. We hence conclude that the internal precision
of our metallicities is 0.1\,dex.

\subsection{Accuracy}

We now consider the accuracy of our metallicities.  The comparison is
based on three external datasets.
The TOPoS sample provides an analysis of the same stars based on higher
  resolution spectra obtained by our group. We have a total of 70
  stars (we excluded stars with upper limits in the high-resolution
  analysis, as well as stars for which \mygi\ did not converge), and
  they are summarised in Table\,\ref{hr}.
There are 88 stars extracted from the PASTEL catalogue
  \citep{pastel} in common with our sample. We excluded the stars
  already present in the TOPoS sample. The data of this sample is
  summarised in Table\,\ref{hrp}.
The low-resolution sample features the stars observed with FORS at VLT
  and GMOS at Gemini by \citet{caffau18}, with resolving power $R\sim
  5000$ and high S/N (typically above 100, 30 stars in common), and by
  \citet{caffau20}, with FORS at VLT, a 
resolving power of $R\sim 2000,$ and
  a high S/N (again above 100, three stars in common). In the
  \citet{caffau18} sample, one star is analysed independently using two
  spectra. The results are very similar, and in the discussion we keep
  both points.

We stress that the comparison with the low-resolution sample is
relevant, since on one hand the spectra have a very high S/N, and on the other hand they have been analysed with a full
multi-element approach; hence, in that analysis Fe and the $\alpha$
elements are measured independently of one another.

In Fig.\,\ref{pl_acc}, we show the literature [Fe/H] as a function of
the metallicity derived from the SDSS spectra by \mygi .  It is
obvious that our abundances are tightly correlated with the external
abundances. In some cases, our abundances are very high compared to the
literature value.  We highlight eight highly deviating
stars in the plot. Three of them are well-known CEMP stars: SDSS\,J029+0238
\citep[][A(C)=8.50]{toposII,toposIII}, SDSS\,J1349-0229
\citep[][A(C)=8.20]{aoki13}, and SDSS 1349-0229
\citep[][A(C)=6.18]{caffau18}. Three other stars (i.e.
SDSS\,J014828+150221, SDSS\,J082511+163459, and SDSS\,J220121+010055)
have a strong G-band that is clearly visible in the SDSS spectrum and
is consistent with a CEMP classification.  We suspect that the
remaining two stars are also carbon-enhanced, although the quality of the
available spectra has not allowed us to confirm that.  The mean
difference (i.e. the external sample minus ours) of the whole
comparison sample (192 stars) is $-0.21$\,dex with a standard
deviation of 0.55\,dex.  If we exclude the eight strongly deviating
stars, reducing the sample to 184 stars, the mean is $-0.12$\,dex with
a standard deviation of 0.35\,dex.

We hence adopted 0.35\,dex as the accuracy of our spectroscopic
metallicities.  This is probably an upper limit to the actual
accuracy, because it does not take into account the errors of the
external metallicities.  There is also an indication that our
metallicity estimates are 0.1--0.2\,dex too high.

\section{Cross-match with  Gaia and down selection\label{sec:gaia}}

As soon as the metallicities from our processing of the SDSS spectra
became available, we began to look at the MDF resulting from this
sample. It was obvious that there was a clear excess of metal-poor
stars, which we suspect is due to the presence of cool white dwarfs,
some of which show \ion{Ca}{ii} K absorption, as suggested by visual
inspection of a few spectra.  We therefore cross-matched our catalogue
with Gaia to use parallaxes to remove white dwarfs from the sample.
To construct the Gaia colour-magnitude diagram, we need to correct the
colours for the effect of reddening.  To be on the same scale as
SDSS, we used the extinction in the $g$ band, $A_g$, provided in the
SDSS catalogue. The reddening on the Gaia colours was estimated in
three steps:
$A_V= A_g/1.161$ ; $E(B-V)=A_V/3.1$;
correct $E(B-V) $ as in Bonifacio et al. (2000), $A_{VC} = 3.1\times E(B-V)_C$;
$(G_{BP}-G_{RP})_0=1.289445\,E(B-V)_C$ ; $G_0 = G - 0.85926\,A_{VC}$.
The extinction coefficients are adopted from the PARSEC web
site\footnote{\href{http://stev.oapd.inaf.it/cgi-bin/cmd}{http://stev.oapd.inaf.it/cgi-bin/cmd}}
\citep{Bressan} and correspond to a G2V star, assuming the extinction
laws by \citet{cardelli} plus \citet{odonnell}, and a
total-to-selective extinction ratio of 3.1.  The cross-match with Gaia
left us 290\,498 spectra for 245\,838 unique stars.
For a large part of the sample, the parallaxes were very poor, leading to
implausibly high luminosities for such stars.

\begin{figure}
\centering
\resizebox{7.5cm}{!}{
\includegraphics{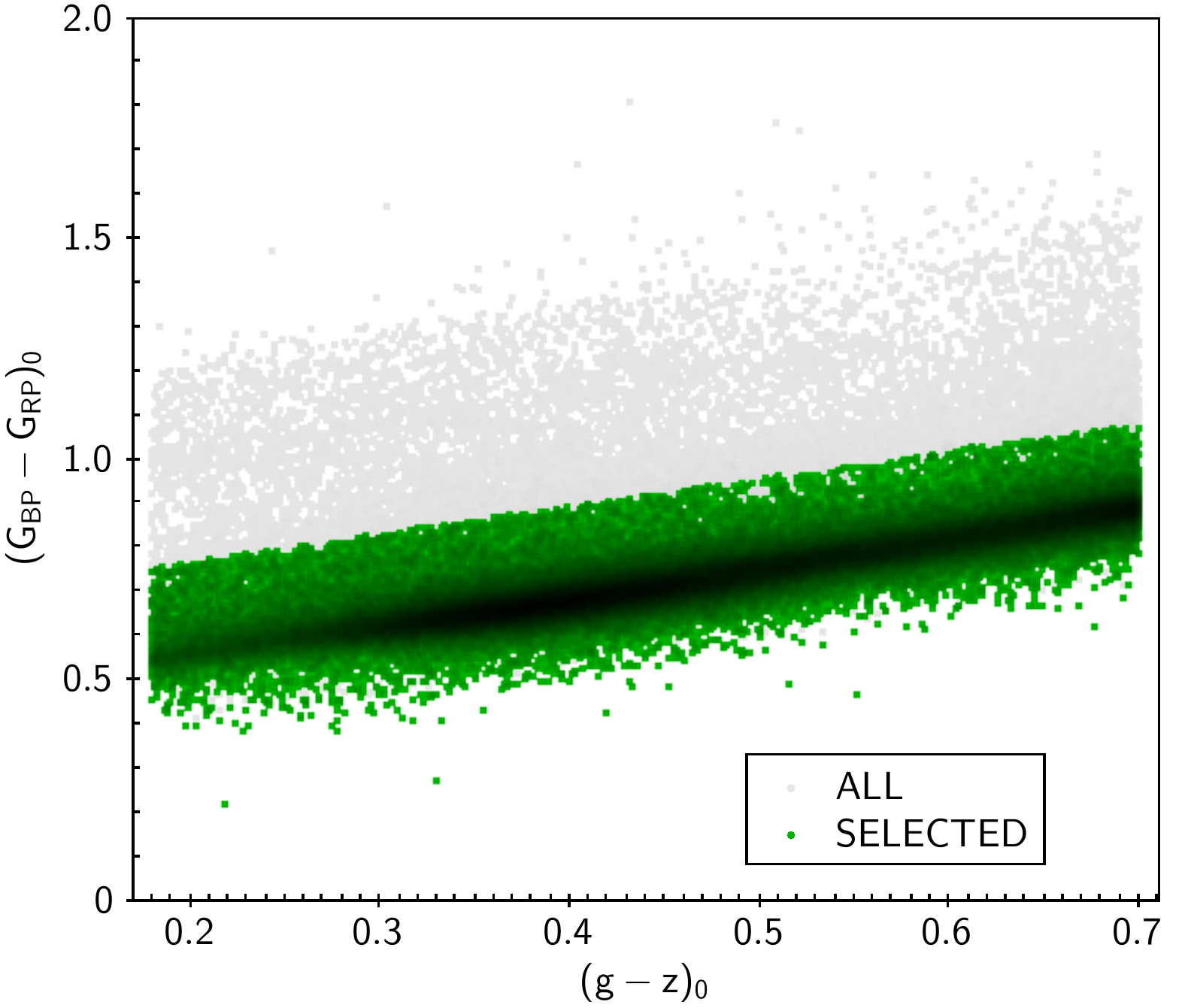}}
\caption{$(g-z)_0$ versus $(G_{BP} - G_{RP})_0$. The full dataset is
  shown as grey dots; the green dots are the selected stars.}
\label{pl_gz_bprp}
\end{figure}

The Gaia photometry, with a typical intrinsic precision of a few mmag,
gives us the opportunity to clean the sample of stars with inaccurate
SDSS photometry.  We started from the $(g-z)_0$ versus $(G_{BP} -
G_{RP})_0$ plane, shown in Fig.\,\ref{pl_gz_bprp}.  We fitted a
straight line\footnote{ $(G_{BP} - G_{RP})_0= 0.43742296099663 +
  0.63260418176651(g-z)_0$; rms = 0.06694397} and selected only the
stars for which the observed $(G_{BP} - G_{RP})_0$ was within $\pm
3\sigma$ of the fitted value.  We assume that the stars that fall out
of this strip have either systematic errors in their $g-z$ colours or
are highly reddened.
Either way, the \teff\ used in the \mygi\ analysis is wrong; therefore,
we discarded these stars.  To exclude all the very-low-luminosity stars,
which we assume to be mostly cool white dwarfs, we also cut on
absolute magnitude, excluding stars with $G_{0,abs} > 7.0$.  Since
many of the stars in our sample have poor parallaxes, we decided to use
the distance estimates of \citet{BJ} to perform this cut. 
{ 
These geometric distances are the result of
a Bayesian estimate, based on the prior of an exponentially decreasing 
space
density. The exponential length scale varies as a function
of Galactic coordinates $(l,b)$. 
For each star, an estimated distance is provided, $r_{est,}$ as well
as a `highest distance' $r_{hi}$  and `lowest distance' $r_{lo}$.
Over the whole
catalogue ($1.33\times 10^9$ stars), the 50th percentile
of the difference $r_{hi}-r_{lo}$ is about 3.6 kpc.
}
This
selection results in 171\,645 spectra for which \mygi\ has converged.
This resulted in 139\,943 unique stars.

\begin{figure}
\centering
\resizebox{7.5cm}{!}{
\includegraphics{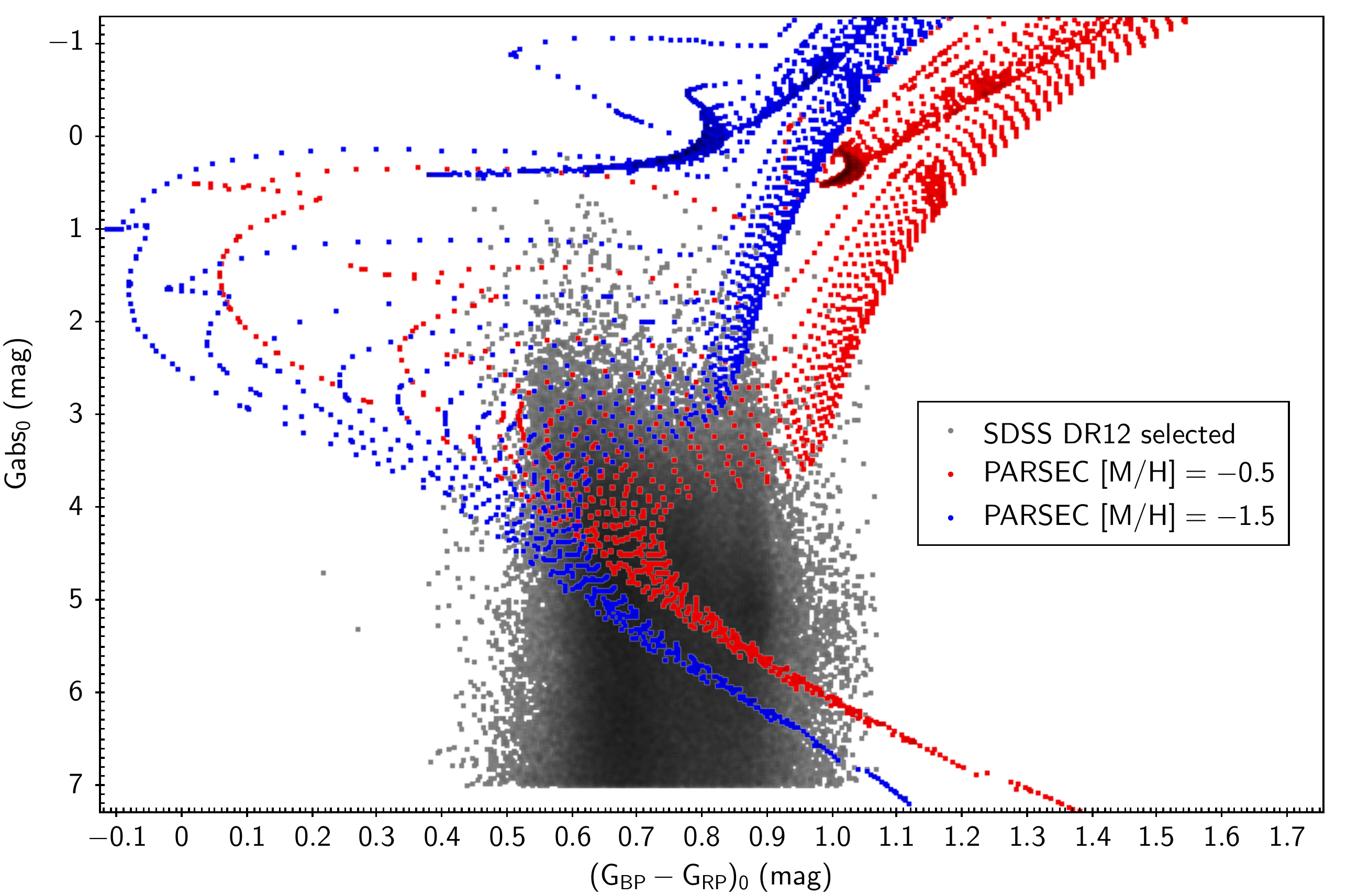}}
\caption{CMD with the selection using Gaia photometry and
  parallaxes.  Two sets of Parsec isochrones \citep{Bressan} of ages
  1--14\,Gyr and metallicities $-0.5$ (red) and $-1.5$ (blue) are
  shown. The absolute magnitudes were computed assuming the
  distances of \citet{BJ}. }
\label{cmd_newsel}
\end{figure}

\begin{figure}
\centering
\resizebox{7.5cm}{!}{
\includegraphics{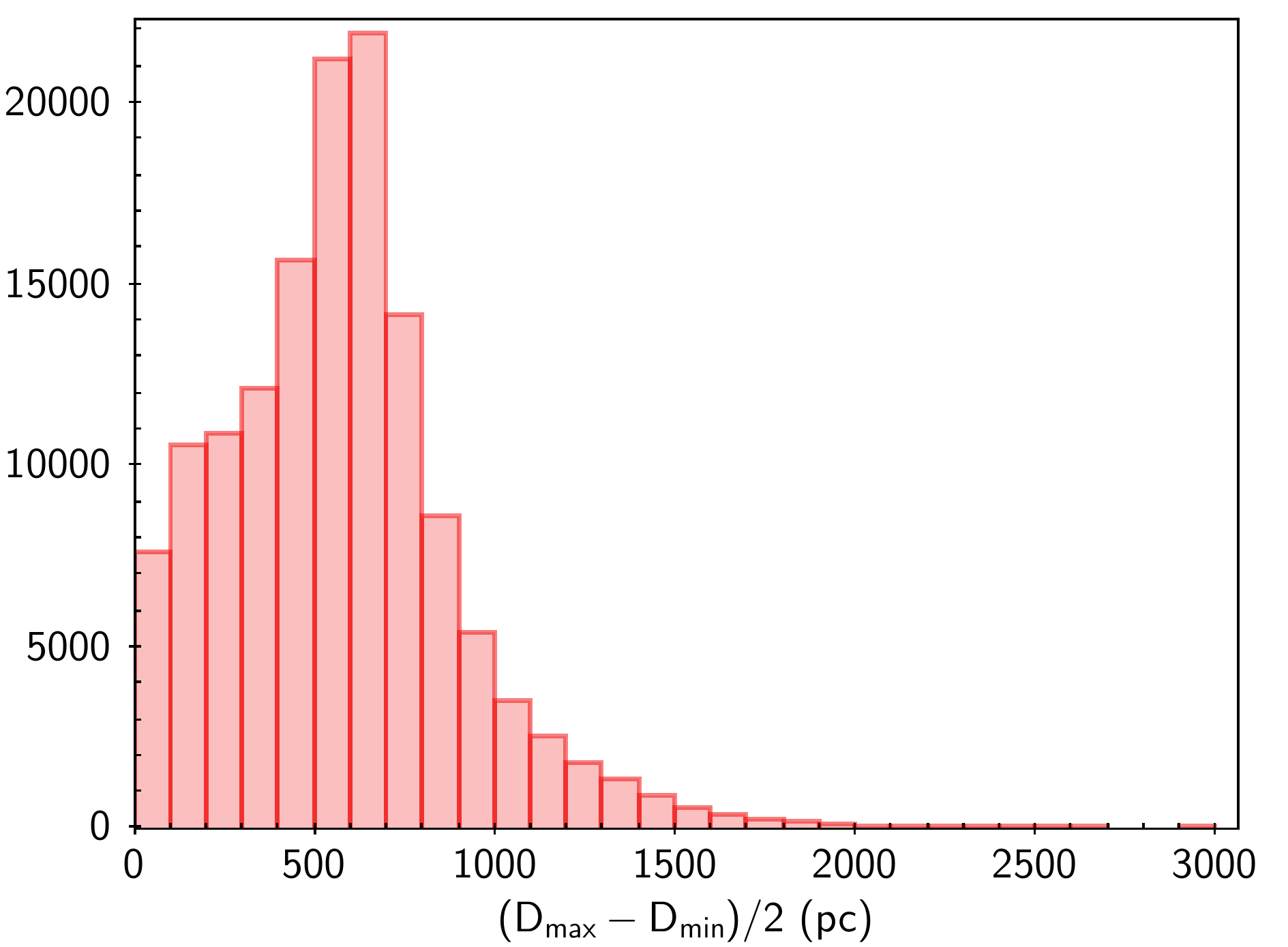}
}
\caption{Histogram of the distance errors in pc, 
from the geometric distances of \citet{BJ}, estimated
as $0.5\times (D_{max} - D_{min})$
}
\label{histo_distance_errors}
\end{figure}

The colour-magnitude diagram (CMD) of the selection is shown in
Fig.\,\ref{cmd_newsel}.  The cleaned sample thus appears to be
compatible with a sample of old TO stars, as is our intention.
{ In Fig. \ref{histo_distance_errors},
we show the histogram of the semi-dispersion of
the geometric distance estimates from \citet{BJ}.
}

\section{Bias in the observed spectroscopic metallicity distribution function}

Our sample consists of 139\,493 unique stars for which we derived
metallicities from at least one SDSS spectrum. For stars with multiple
spectra, each spectrum was independently analysed, and then the derived
metallicities were averaged (see Sect. \ref{metprecision}).  Our
spectroscopic sample contains stars that have been observed in four
different SDSS programs, comprising 39 different surveys. It seems
arduous to extract the bias in our sample starting from the selection
criteria of the different surveys convoluted with our own selection
criteria. In Table \ref{types}, we assembled the programmes and the
types of our sample of spectra, with the numbers for each.  We note that
6\,425 spectra have a blank type, for 6221 targets the type is 'not
available' ({\tt NA}), and for 20\,973 it is {\tt STAR}. The largest
fraction of our targets are {\tt NONLEGACY}.  We were unable to
find any reference to this type of target in the SDSS documentation,
and so we may guess that these are essentially `fill fibres' to complete a
plate, with no particular selection criterion.  The sample is in fact
dominated by these targets and spectro-photometric standards.  We
therefore employed a different route to estimate the bias in our
sample.
\begin{table}
\caption{
Target types and SDSS programs for which the spectra have been observed
}
\label{types}
\centering
\begin{tabular}{lr}
\hline
Type & \# of spectra \\\hline
~~~~~~~      &   6425\\ 
GALAXY       &    736\\
NA           &   6221\\
NONLEGACY    &  69917\\
QA           &   1308\\ 
QSO          &   5202\\
REDDEN\_STD   &  10106\\  
ROSAT\_D      &    601\\
SERENDIPIT   &     31\\
SPECTROPHO   &  47930\\
STANDARD     &      4\\
STAR         &  20973\\
STAR\_BHB     &   3032 \\
STAR\_WHITE   &    159\\\hline
Programme & \# of spectra \\\hline
boss  &45250\\
sdss  &42543\\
segue1&58298\\
segue2&26554\\
\hline
\end{tabular}
\end{table}

\subsection{Photometric metallicities}

We began by determining metallicities using a photometric index that
is metallicity sensitive. The obvious choice would be $u-g$; however,
we preferred to use a reddening-free index that can be formed combining
$u-g$ and $g-z$:
\begin{equation}
p = (u-g) - 0.5885914\,(g-z).
\end{equation}

{ 
The choice of the colours to form 
the $p$ index is dictated by the need to
have a metallicity-sensitive colour,
the fact that $u-g$ is the most metallicity-sensitive colour in $ugriz$ photometry,
and the need to factor out the temperature
sensitivity of the same index. The 
choice of $g-z$ has been done
to be consistent with the colour chosen to 
determine effective temperatures. 
The coefficient  in front of $g-z$ is nothing other
than the ratio of the two reddening
coefficients as provided by \citet{Schlegel}
in order to make $p$ reddening free.
In fact, $p$ is nothing other than the $ugz$
equivalent of the $Q$ index introduced
by \citet{JM53} for $UBV$ photometry.
}

\begin{figure}
\centering
\resizebox{7.5cm}{!}{
\includegraphics{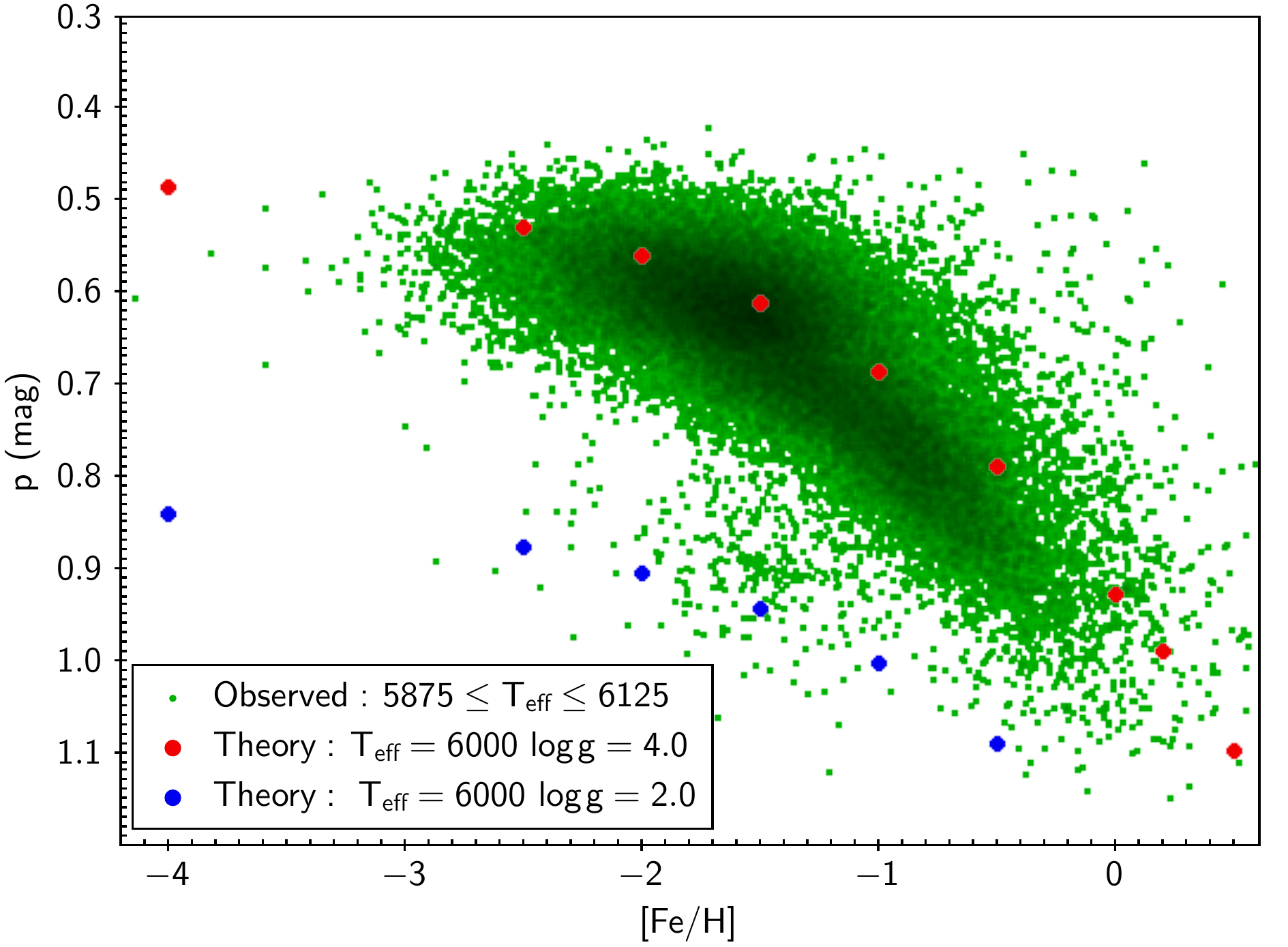}}
\caption{$p$ index as a function of metallicity for stars with
  5875\,K $\le$\teff $\le 6125$\,K compared with theoretical
  predictions for $\log g = 4.0$ (red) and $\log g = 2.0$ (blue),
  derived from the \citet{CK03} grid of ATLAS 9 model atmospheres.
{ For the observed stars, we show the spectroscopic metallicities.}}
\label{p_met_t600}
\end{figure}

The $p$ index varies with effective temperature, surface gravity, and
metallicity, as illustrated in Fig.\,\ref{p_met_t600}, 
{ where we show the spectroscopic metallicities on the x-axis}. It is clear
that a spread in $p$ at any given metallicity remains, even by fixing
the effective temperature of the sample. This is because $p$ is also
sensitive to $\log g$, and the set of a few stars that are around
metallicity $-1.5$ and $p\approx 0.9$ also contains horizontal-branch
stars. In order to obtain a metallicity, we adopted a very simple 
approach: we derived \teff\ from the $(g-z)_0$ colour, and once the
\teff\ is known, we derived the surface gravity using the parallax and
the Stefan-Boltzmann equation, assuming a mass of 0.8M\sun.  Once
\teff\ and $\log g$ were known, we only needed to interpolate in a grid
of theoretical values to obtain a metallicity from the observed $p$
index.

We computed the theoretical $p$ index for the \citet{CK03} grid of
ATLAS 9 model atmospheres.  The procedure needs to be iterated, since
to determine the surface gravity one needs to know the bolometric
magnitude. The latter can be derived from the Gaia $G_0$ magnitude
and a theoretical bolometric correction that we computed for all the
\citet{CK03} grid. To do this, one needs to know the metallicity. In
this sense, the procedure is iterative: one starts from a guess of the
metallicity to determine the bolometric correction and then computes the
metallicity. At that point, with the new metallicity, one derives a new
bolometric correction and iterates the procedure. In the vast majority
of cases, a single iteration was sufficient. Bolometric corrections for
the $G$ magnitude are small for TO stars anyway.  In order to accelerate the
procedure, we did not interpolate in the grid of theoretical $p$
values for the current \teff\ and $\log g$, but we simply used the grid
values that are nearest.
In order to be able to use the method, we need parallaxes for all the
stars. For the majority of the stars in our spectroscopic and
photometric samples, the Gaia parallaxes are too uncertain to be
used. Therefore, to derive the metallicities we used the distance
estimates provided by \citet{BJ}, which were derived from Gaia parallaxes
using a Bayesian estimate and a prior of exponentially decreasing
stellar density.
{ The procedure to determine the metallicities from the $p$ index is 
explained in detail in Appendix \ref{appc}.}

\begin{figure}
\centering
\resizebox{7.5cm}{!}{
\includegraphics{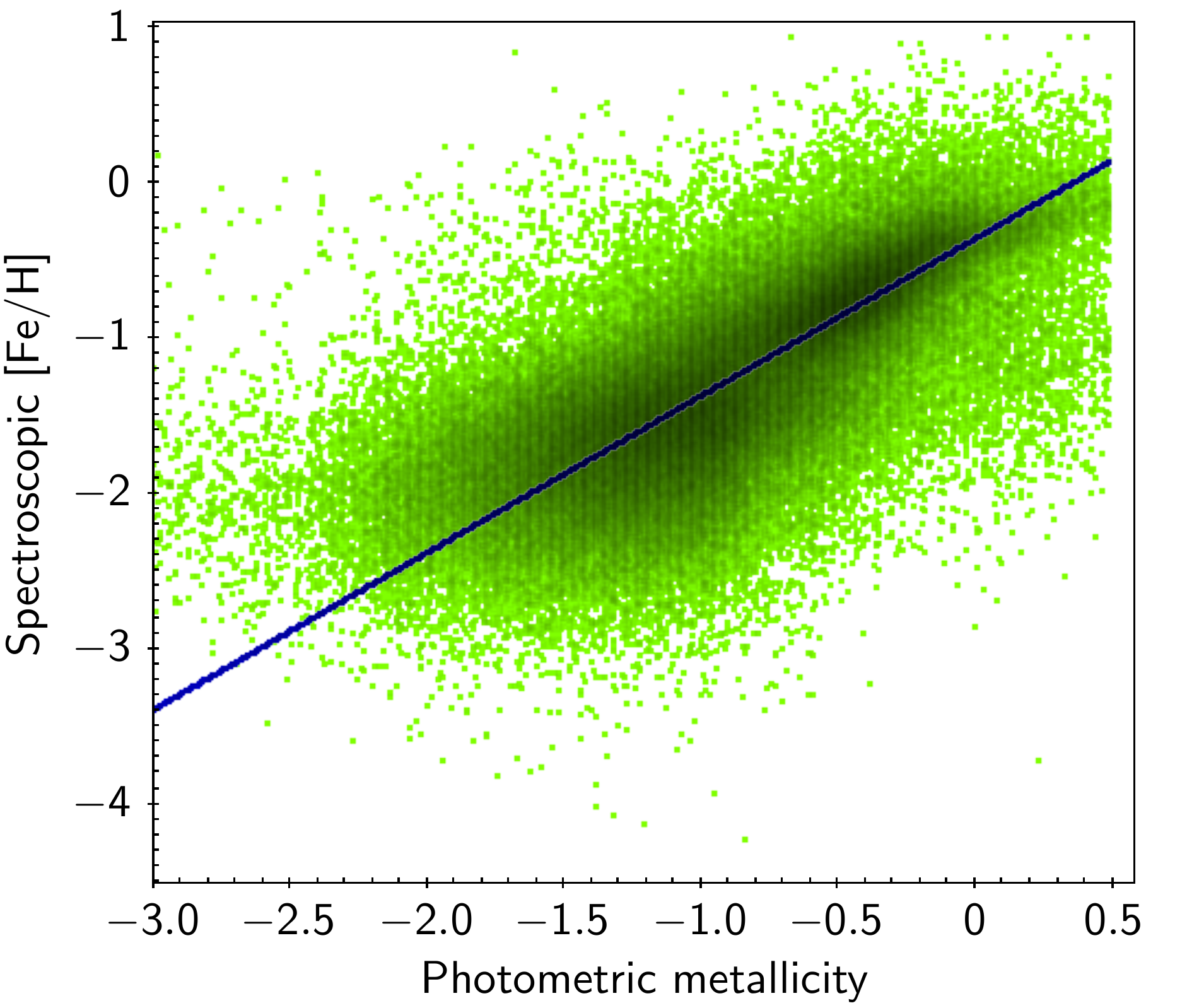}}
\caption{Spectroscopic metallicity as a function of the
  metallicity derived from the $p$ index. The blue line is a linear
  fit we made to the sub-set of stars along the highest density ridge.
}
\label{pmet_phot_spec}
\end{figure}

\begin{figure}
\centering
\resizebox{7.5cm}{!}{
\includegraphics{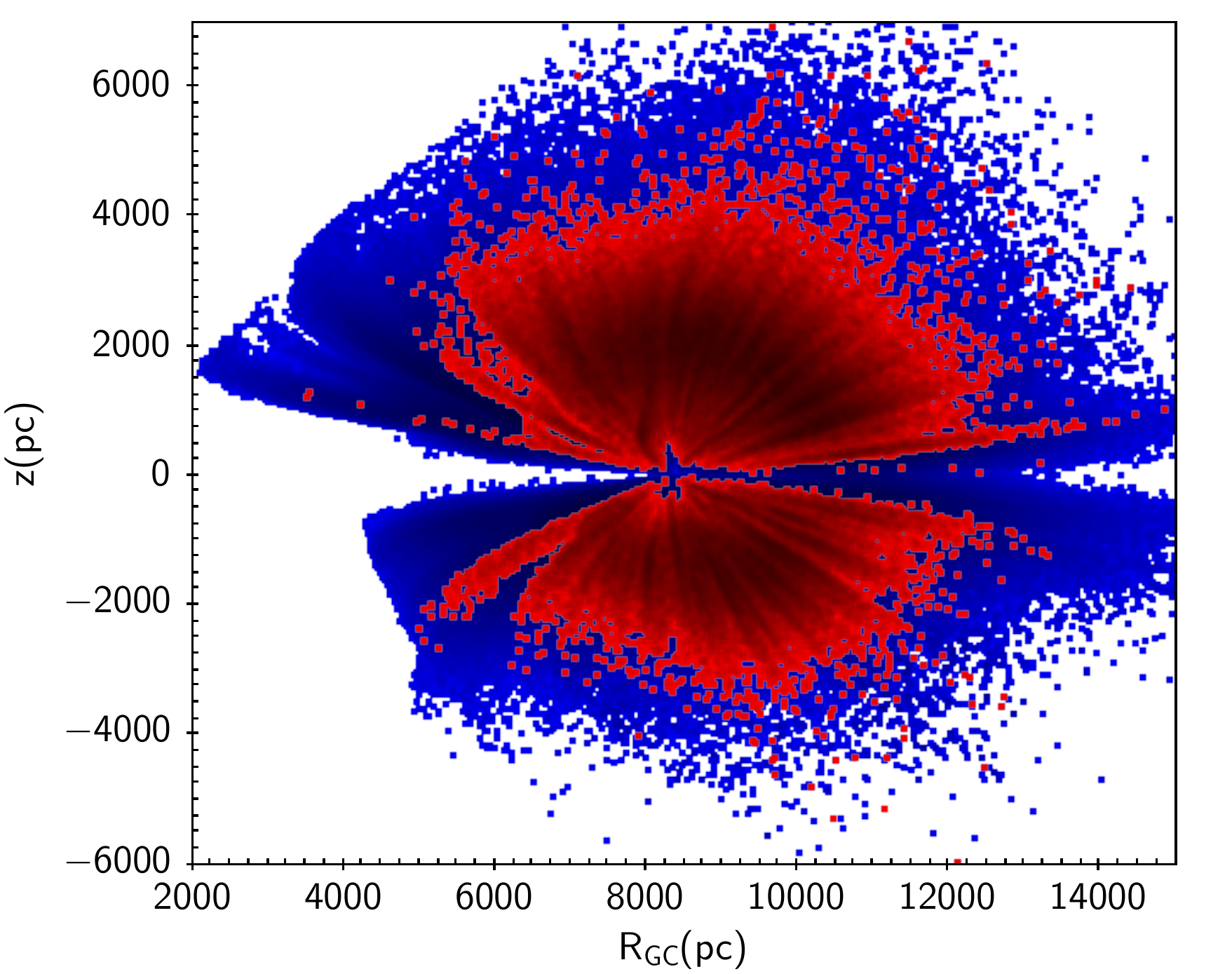}}
\caption{Spectroscopic (red) and photometric (blue) samples in
  distance from the Galactic centre and height above the Galactic
  plane.}
\label{rgc_z}
\end{figure}

\begin{figure}
\centering
\resizebox{7.5cm}{!}{
\includegraphics{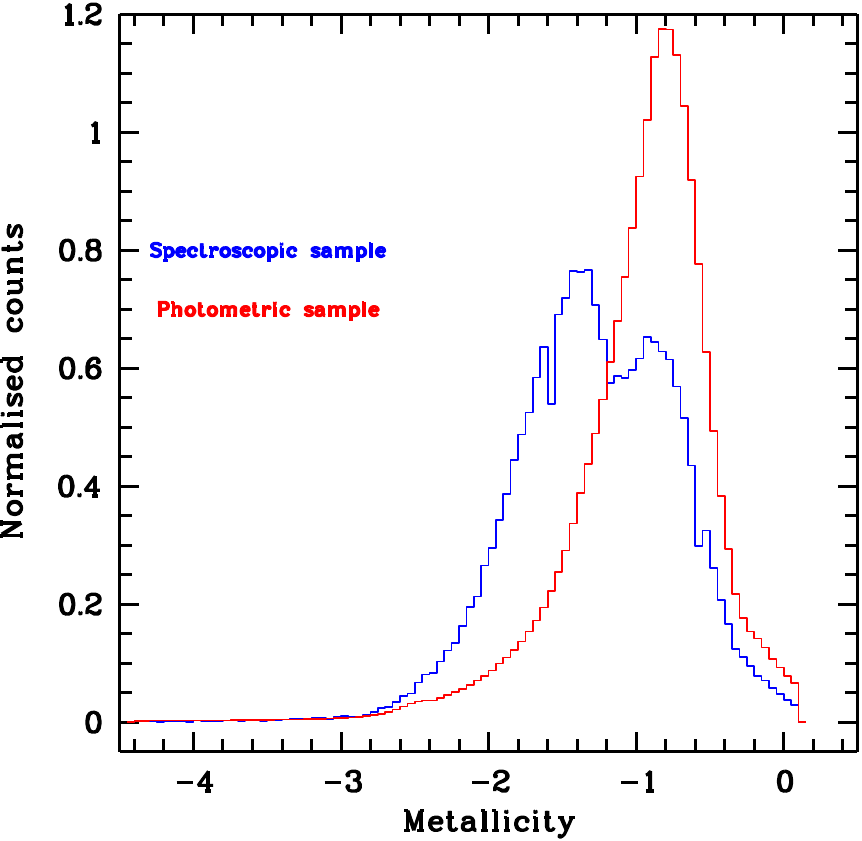}}
\caption{Distribution of photometric metallicities for the photometric
  and spectroscopic samples. Each sample has been normalised so
  that the area under the histogram is equal to one.  }
\label{hist_ph_s}
\end{figure}

In Fig.\,\ref{pmet_phot_spec}, we show the comparison of photometric
and spectroscopic metallicities. Aside from the large scatter, which we
attribute mainly to the large uncertainty in the photometric
metallicity estimate, the two variables are tightly correlated. We{ interactively selected a
sub-set of 118\,158  stars that lie along the highest density ridge} and
fitted a straight line (shown in blue) to derive the spectroscopic
metallicity from the photometric metallicity.

\begin{figure}
\centering
\resizebox{7.5cm}{!}{
\includegraphics{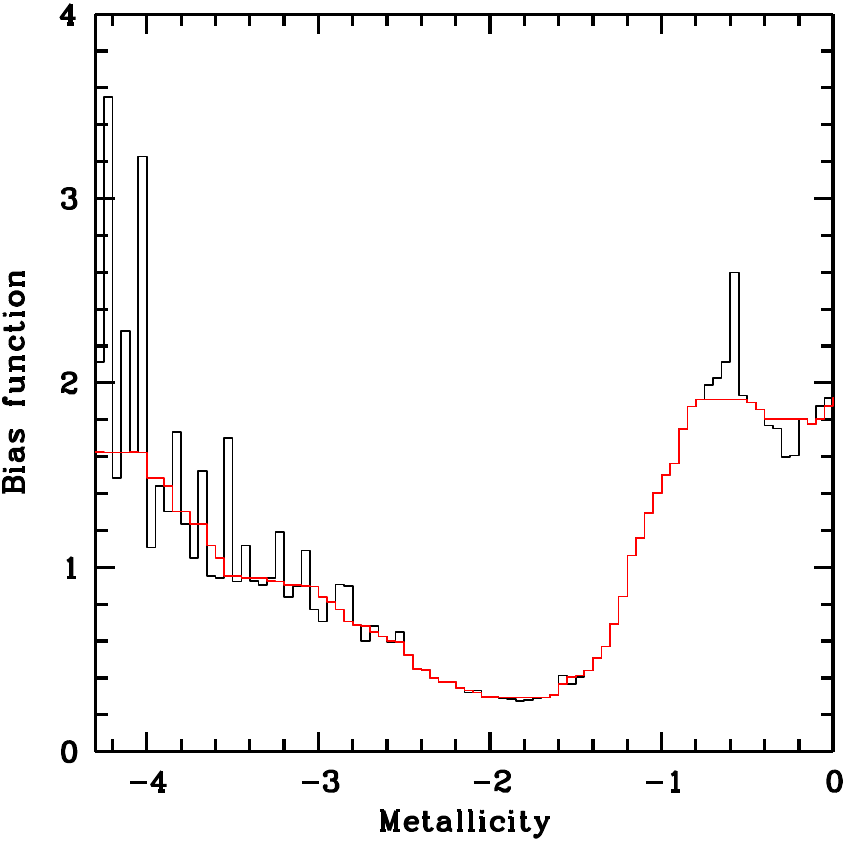}}
\caption{Bias function (black line) and {smoothed} bias function (red line)}
\label{bias}
\end{figure}

\begin{figure}
\centering
\resizebox{7.5cm}{!}{
\includegraphics{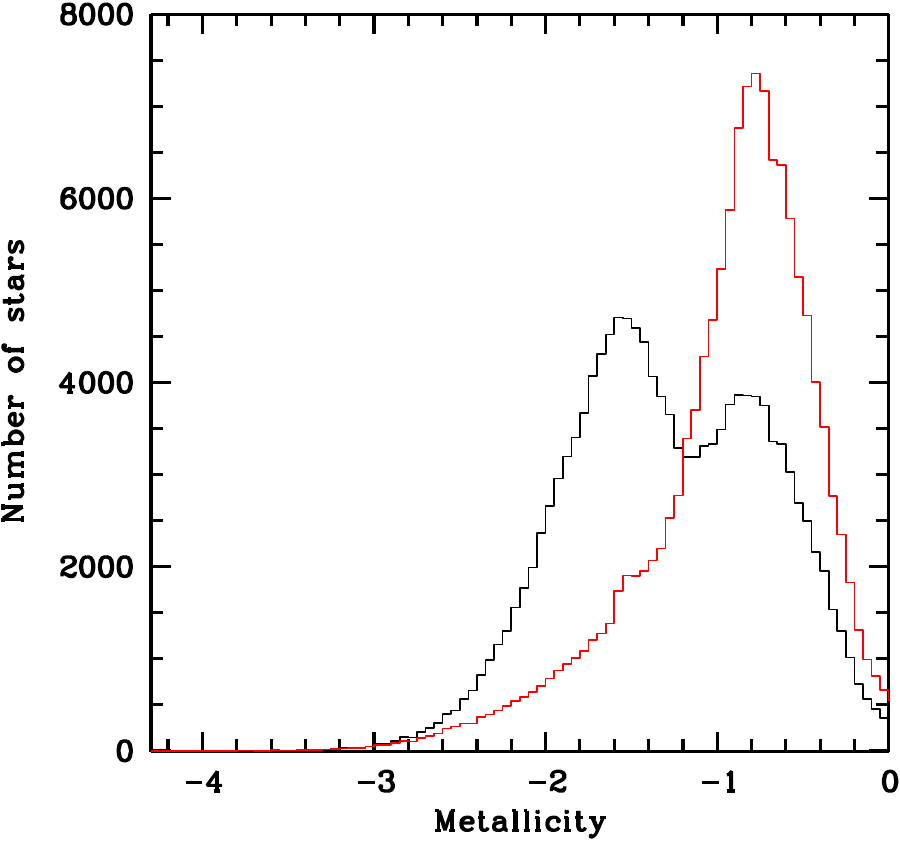}}
\caption{Raw metallicity distribution function (black line) and
  bias-corrected (red line).}
\label{pl_histo_c}
\end{figure}

\begin{figure}
\centering
\resizebox{7.5cm}{!}{
\includegraphics{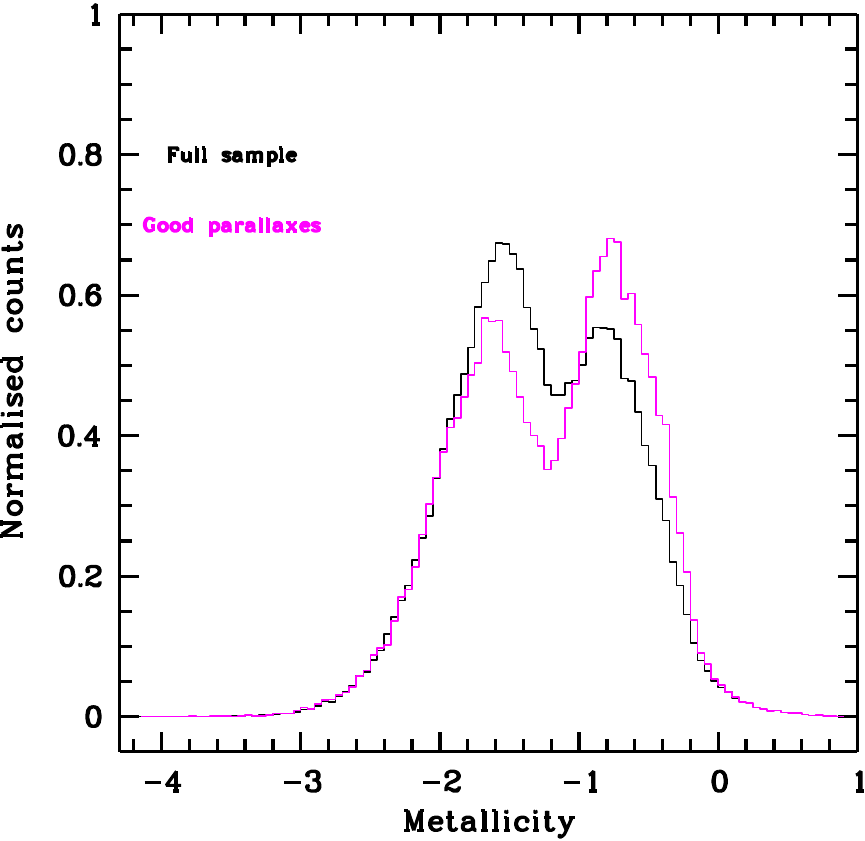}}
\caption{Normalised metallicity distribution of our full spectroscopic
  sample (black) compared to the subset with good parallaxes
  (violet). The normalisation is such that the area under each
  histogram is equal to one.}
\label{hist_good}
\end{figure}

\begin{figure}
\centering
\resizebox{7.5cm}{!}{
\includegraphics{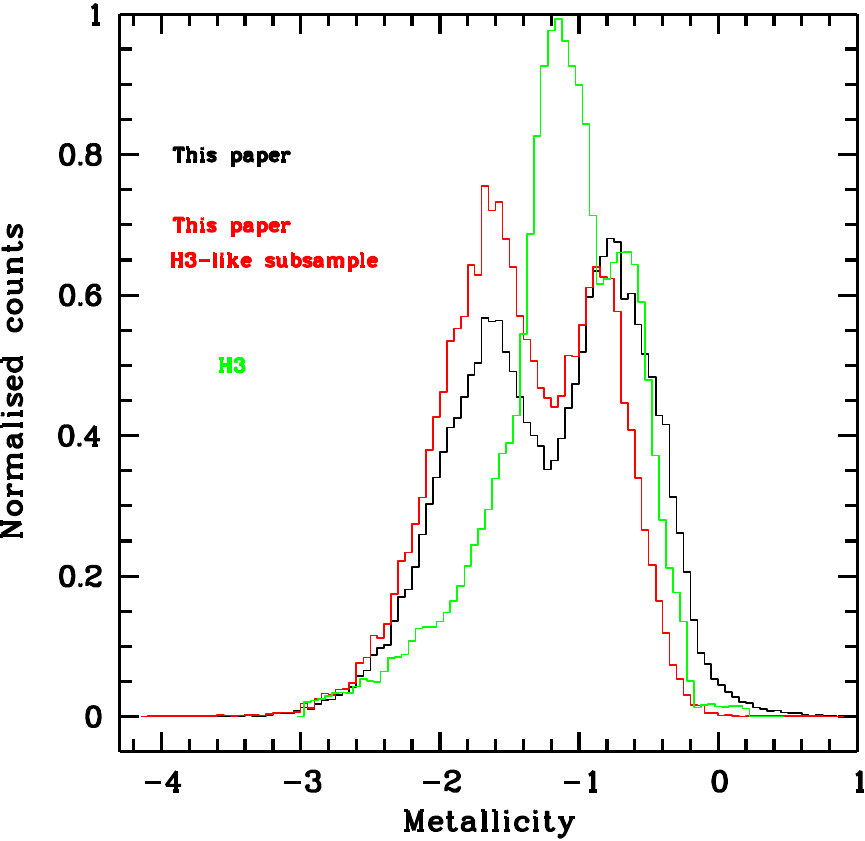}}
\caption{Normalised MDFs for our sample with good parallaxes (black
  line), the sub-sample selected according to the criteria of H3 (red
  line) and the H3 \citep[][green]{Naidu20}.}
\label{pl_h3_mygi}
\end{figure}

\begin{figure}
\centering
\resizebox{7.5cm}{!}{
\includegraphics{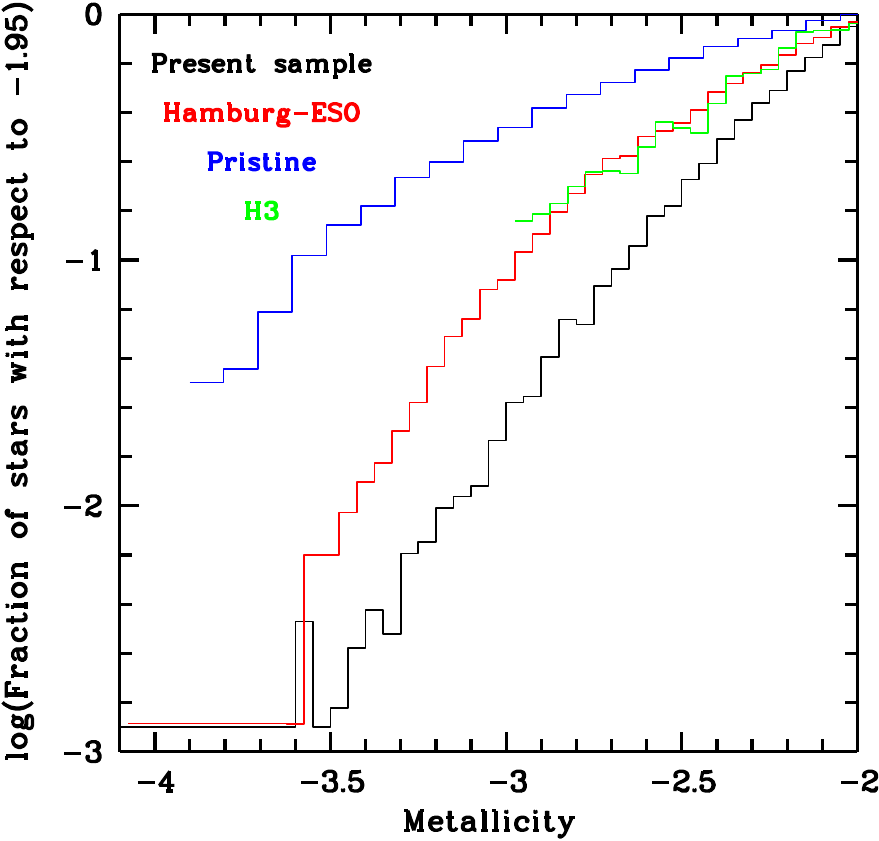}}
\caption{Metal-weak tail of the raw metallicity distribution function
  (black line), normalised at $-1.95$, compared with the MDFs of the
  Hamburg/ESO survey \citep[][red]{Schoerck}, the Pristine survey
  \citep[][blue]{Youakim}, and the H3 survey
  \citep[][green]{Naidu20}.}
\label{pl_raw_mdf_n2}
\end{figure}

\begin{figure}
\centering
\resizebox{7.5cm}{!}{
\includegraphics{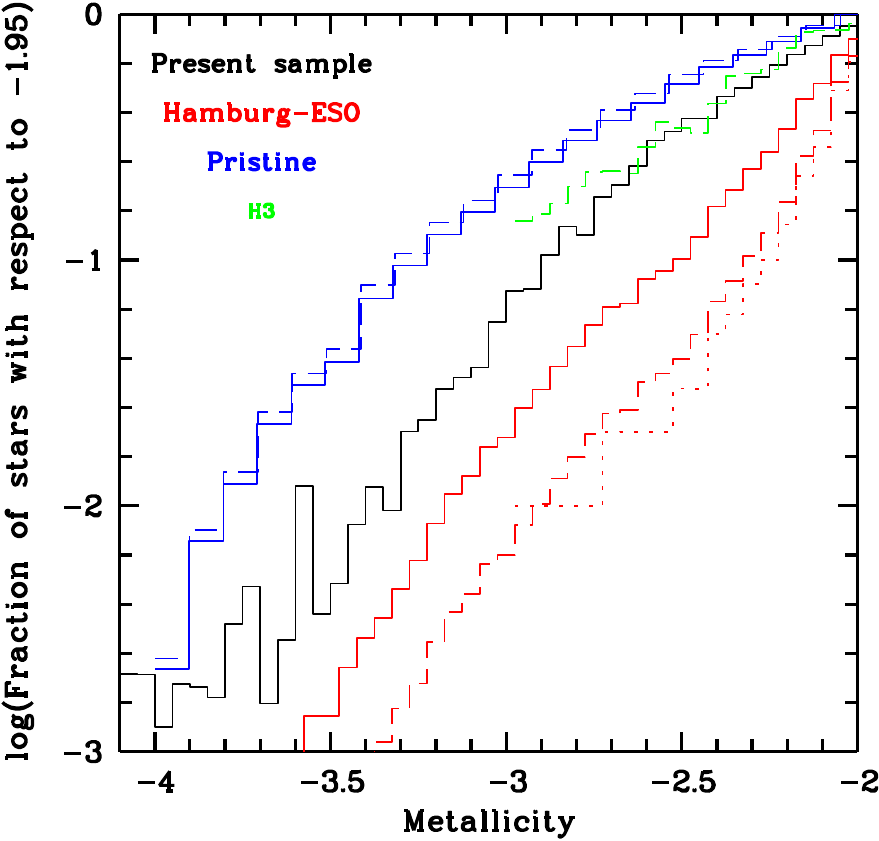}}
\caption{Metal-weak tail of the corrected metallicity distribution
  function (black line), normalised at $\mathrm{[Fe/H]}=-1.95$,
  compared with the corrected Hamburg/ESO \citep[][red]{Schoerck} and
  the corrected Pristine \citep[][blue]{Youakim} MDFs.}
\label{pl_corr_mdf_n2}
\end{figure}

By construction, these `adjusted' photometric metallicities{ 
(i.e. not those derived directly from $p$ using the procedure in Appendix
\ref{appc}, but those derived from the latter and the straight line fit
shown in Fig.\,\ref{pmet_phot_spec})} have a
distribution that is similar to that of the spectroscopic
metallicities.

\subsection{Comparison with a larger photometric sample\label{sec:comp_phot}}

In order to estimate the bias in our spectroscopic sample, we make the
bold assumption that a sample selected from the SDSS~DR12 photometry,
using the same colour and absolute magnitude cuts that have defined
our spectroscopic sample, is bias-free. Having selected the sample, we
applied to it our photometric calibration.{ The distances
were adopted from the \citet{BJ} catalogue,} and we discarded about one million
stars whose colours fall outside our synthetic grid. The final
sample, which we refer to in the following as `photometric',
consists of 24\,037\,008 unique stars.

In Fig. \ref{rgc_z}, we show the two samples in the diagram of
height above the Galactic plane and distance from the Galactic centre.
We adopted a distance of 8.43\,kpc for the Galactic centre \citep{Reid14}
and a distance of the Sun above the Galactic plane of 27\,pc \citep{Chen01}.
The photometric sample reaches higher values of $|z|$ and
larger distances from the Sun. This is a result of the fact that the photometric
sample has many more faint stars than the spectroscopic sample. 
We computed the bias using the full sample (see Sect.~\ref{sec:bias})
and also using a sub-set of the photometric sample that covers
the range in $|z|$ and $R_{GC}$ occupied by the spectroscopic sample more closely.
The difference in the derived bias is negligible.

We can now compare the two metallicity distributions.  In
Fig.\,\ref{hist_ph_s}, the distributions of the photometric
metallicities of the two samples are compared.  It is clear that the
spectroscopic sample is boosted in metal-poor stars. An obvious reason
for this is the large number of spectrophotometric standards that are
part of the sample. These stars are selected to have colours
consistent with metal-poor F dwarfs to allow a reliable flux
calibration of the spectra on the same plate. We suspect that many of
the {\tt NONLEGACY} targets were selected according to similar criteria.
These two contributions alone could explain the boost of metal-poor
stars in the spectroscopic sample.

\subsection{Correcting for the bias\label{sec:bias}}

A straightforward way to correct our sample for the selection bias is
to take the ratio of the two histograms in Fig.\,\ref{hist_ph_s}, (photometric sample)/(spectroscopic sample), to derive a
`bias function'. This function is shown in Fig.\,\ref{bias} and
relies on { photometric} metallicities in { both} samples.  In
the same plot, we also show a smoothed version of the bias function
(median filter with a 0.25\,dex kernel) that avoids the oscillations
seen at the extremes of the metallicity range that are due to the
small number of stars contained in each bin in either of the two
histogram functions. We can then produce a bias-corrected version of
the histogram of our { spectroscopic} metallicities, simply by
multiplying the observed histogram with this bias function.  This
rests on the assumption that the bias in { spectroscopic}
metallicities is the same as that in { photometric}
metallicities. Or to put it another way, if we had an SDSS spectrum
for each of the stars in the photometric sample and derived a
spectroscopic metallicity for it, the bias derived dividing the
histogram of these { spectroscopic} metallicities by that of the
{ spectroscopic} metallicities of the current spectroscopic sample
would be the same as that shown in Fig.\,\ref{bias}.

The result of this exercise is shown in Fig.\,\ref{pl_histo_c}, where the
observed and bias-corrected MDFs are shown for our spectroscopic sample.
Two features in the bias-corrected MDF are obvious: {\em i)} the bi-modality
disappears and the metal-poor peak is now detectable only as a change in the slope
of the MDF; and {\em ii)} the slope of the MDF from a metallicity  $\sim -1.6$ is
much more gentle than in the observed MDF.

\section{The  MDF\label{sec:mdf}}

Previous investigations of the MDF used different techniques. To keep
the comparison to our results manageable, we selected just three MDFs: \citet[][Hamburg/ESO Survey, hereafter
  HES]{Schoerck},\ \citet[][H3 Survey]{Naidu20}, and \citet[][Pristine
  Survey]{Youakim}.  The first two are based on spectra, while the
third uses a photometric metallicity estimate.  Our MDF is provided in
Table\,\ref{tabmdf}.

The HES MDF is derived from a sample of metal-weak candidates
selected from the objective prism spectra \citep[see][]{Christlieb08},
then observed at low resolution ($R\sim 2000$) with several telescopes
and spectrographs, to provide a sample of 1638 unique stars.

The H3 sample presented in \citet{Naidu20} is selected with the following
conditions: {\em i) } $15 < r < 18$; $\varpi - 2\times \sigma_\varpi <
0.5$; $|b| > 40^\circ$.  \citet{Naidu20} claim that this sample is
unbiased with respect to metallicity.
{
The stars selected by H3 
are observed with the multi-fibre Hectoechelle spectrograph
\citep{Hectoechelle} on the 6m MMT telescope \citep{H3}.
The spectra cover the range 513 --530\,nm
at a resolving power R$\sim 23\,000$. Using the {\tt MINESweeper} code
\citep{Cargile}, an analysis of the spectra combined with photometry
provides
radial velocities, spectrophotometric distances, 
and abundances ([Fe/H] and [$\alpha$/Fe]). 
}

The Pristine sample is aimed at selecting the tip of the main-sequence
turn-off of the old populations of the Galaxy using SDSS photometry.
A magnitude cut was applied to select stars with $19 < g_0 < 20$ and
colour $0.15 < (g-i) < 0.4$, combined with the exclusion of a region
in the $(u-g)_0$ versus $(g-i)_0$ plane that is dominated by young
stars \citep[see Fig. 3 of][]{Youakim}. This sample amounts to
about 80\,000 stars.  A bias correction of the sample is proposed to
correct for the selection effects and for the photometric errors.

We did not compare to the MDF of \citet{Allende14}, because its
metal-weak tail is essentially identical to ours, as is expected,
since there is a substantial overlap between the two sets of spectra,
although the spectra were analysed in different way.  No bias
correction was attempted in \citet{Allende14}.  With respect to our
sample, the \citet{Allende14} sample lacks the metal-rich peak, and
this is just a matter of selection criteria.

\subsection{Comparison with H3}

We carried out a more detailed comparison with H3, since it is the one that is
more easily directly comparable to our results. Both are based on
spectroscopy, and an MDF is available up to super-solar metallicities.

Here, and in Section \ref{dynamics}, we need to define sub-samples that
make use of parallaxes. For this reason, we need to define a sub-set
with `good' parallaxes.  We define this sample as stars for which
$\varpi > 3\times\Delta\varpi$, which leaves us with a sample of
57\,386 unique stars. In terms of metallicity distribution, this
sub-sample favours the metal-rich component, so that now the most
prominent peak is the metal-rich one, as shown in Fig.\,\ref{hist_good}.
On the other hand, the metal-weak tail of the two distributions is
pretty much the same. In the following, we refer to this
sub-sample as the good parallax sample.  To have more meaningful
comparison between the H3 MDF and our sample, we selected a sub-sample of our
good parallax sample, which fulfils the same selection criteria as
\citet{Naidu20}. This is composed of 28\,165 stars, which is 49\,\% of
our good parallax sample. We refer to this sub-sample as
the H3-like sample. In Fig.\,\ref{pl_h3_mygi}, we compare the normalised
MDFs of our good parallax and of our H3-like samples to that of
\citet{Naidu20}.

From this comparison four facts are obvious:
Our sample is boosted in metal-poor stars with respect to an
  unbiased sample, confirming what was deduced from the comparison with our
  photometric sample.
A H3-like selection, although it does not make any explicit
  reference to metallicity, is likely to introduce a bias in
  metallicity. This can be inferred by the comparison of our good
  parallax and H3-like samples. { It is, however, obvious that this
  bias is minor, especially in the low-metallicity tail, and we ignore it in the following}.
The H3 MDF has a much more gentle slope of the metal-weak tail
  than our sample.
The metal-rich peak in our{ full }sample seems very similar to that in
  the H3 sample.

\subsection{The metal-weak tail of the MDF}

While the metal-rich portion of the MDF is the result of the complex
chemical evolution of the Galaxy, we may assume that its metal-weak
tail is much simpler, since many of the stars are the descendants of
only a few generations of stars beyond the first one.  For more than a decade, theoretical models have shown that the
properties of the first stars affect the metal-weak tail of the MDF
\citep[e.g.][]{HF01,P03,O03,K06,T06,Salvadori07}.
Furthermore, this should hold true even if this metal-weak tail
contains stars that have been formed in Milky Way satellites that
later merged to form the present-day Milky Way
\citep[e.g.][]{Salvadori15}.

To study the metal-weak tail of the MDF, \citet{Schoerck} normalised
their distributions at metallicity $-1.95$.  This normalised
distribution provides, for any lower metallicity, the fraction of
stars in that metallicity bin, with respect to those in the bin
centred at $-1.95$. We use the same approach, which also allows
us to directly compare our results with theirs.

In Fig.\,\ref{pl_raw_mdf_n2}, we compare the normalised metal-weak
tail of the MDF, as observed, for {\em i)} our { whole }sample; {\em ii)} the
Hamburg/ESO sample \citep{Schoerck}; {\em iii)} the Pristine sample
\citep{Youakim}; and {\em iv)} the H3 sample \citep{Naidu20}. The H3
Survey adopts a target selection that is based only on apparent
magnitudes, parallaxes, and Galactic latitude, thus they claim that
their sample is unbiased with respect to metallicity. It is
surprising that the metal-weak tail of their MDF is essentially
identical to that of the Hamburg-ESO Survey down to metallicity $-3.0$,
since the Hamburg/ESO Survey has been considered to be biased in
favour of metal-poor stars \citep{Schoerck}.  The abrupt drop of the
H3 MDF at $-3.0$ is not easy to understand, but it may be due to the fact
that the size of their sample is too small to adequately sample the
populations of lower metallicity.

The selection function for the four samples is clearly
different. \citet{Schoerck} provide three possible bias corrections
for their observed MDF, while \citet{Youakim} provides two.  In
Fig.\,\ref{pl_corr_mdf_n2}, we show the comparison with all the
corrected MDFs. What is obvious is that while in the raw
distributions ours is always {\em below} the Hamburg/ESO, once bias
corrections are taken into account ours is {\em above} the
Hamburg-ESO. As noticed in Sect.\,\ref{sec:bias}, this is due to the
fact that in our corrected MDF, the slope is much more gentle than in
our raw MDF; as a consequence, once we normalise at $-1.95$, the
corrected MDF lies above the raw MDF. The reverse is true if we look
at the non-normalised MDF, as shown in Fig.\,\ref{pl_histo_c}.

\begin{table}
\caption{Raw and corrected metallicity distribution functions
for the  139\,493 unique stars of our sample.\label{tabmdf}}
\begin{tabular}{crr|crr}
\hline 
[M/H]    &  raw    &  corrected & [M/H]    &  raw    &  corrected     \\
\hline
$ -4.43 $ &     0   &    0&$ -2.12 $ &  1771   &  586\\
$ -4.38 $ &     0   &    0&$ -2.08 $ &  1990   &  640\\
$ -4.32 $ &     0   &    0&$ -2.03 $ &  2368   &  703\\
$ -4.28 $ &     0   &    0&$ -1.98 $ &  2656   &  787\\
$ -4.22 $ &     1   &    2&$ -1.92 $ &  2955   &  873\\
$ -4.18 $ &     0   &    0&$ -1.88 $ &  3196   &  944\\
$ -4.12 $ &     1   &    2&$ -1.83 $ &  3403   & 1005\\
$ -4.07 $ &     1   &    2&$ -1.77 $ &  3666   & 1083\\
$ -4.03 $ &     1   &    2&$ -1.73 $ &  4069   & 1202\\
$ -3.97 $ &     0   &    0&$ -1.67 $ &  4311   & 1273\\
$ -3.92 $ &     1   &    1&$ -1.62 $ &  4520   & 1383\\
$ -3.88 $ &     1   &    1&$ -1.58 $ &  4703   & 1733\\
$ -3.83 $ &     1   &    1&$ -1.52 $ &  4696   & 1902\\
$ -3.78 $ &     2   &    3&$ -1.48 $ &  4592   & 1896\\
$ -3.72 $ &     3   &    4&$ -1.42 $ &  4441   & 1957\\
$ -3.67 $ &     1   &    1&$ -1.38 $ &  4063   & 2069\\
$ -3.62 $ &     2   &    2&$ -1.33 $ &  3849   & 2198\\
$ -3.58 $ &     9   &    9&$ -1.27 $ &  3653   & 2529\\
$ -3.53 $ &     3   &    3&$ -1.23 $ &  3289   & 2772\\
$ -3.47 $ &     4   &    4&$ -1.17 $ &  3191   & 3393\\
$ -3.42 $ &     7   &    7&$ -1.12 $ &  3189   & 3698\\
$ -3.38 $ &    10   &    9&$ -1.08 $ &  3310   & 4283\\
$ -3.33 $ &     8   &    8&$ -1.02 $ &  3334   & 4678\\
$ -3.28 $ &    17   &   16&$ -0.98 $ &  3487   & 5230\\
$ -3.22 $ &    19   &   18&$ -0.93 $ &  3760   & 5874\\
$ -3.17 $ &    26   &   23&$ -0.88 $ &  3862   & 6761\\
$ -3.12 $ &    29   &   26&$ -0.82 $ &  3856   & 7215\\
$ -3.08 $ &    32   &   29&$ -0.77 $ &  3849   & 7354\\
$ -3.03 $ &    49   &   44&$ -0.73 $ &  3748   & 7161\\
$ -2.97 $ &    70   &   59&$ -0.68 $ &  3357   & 6414\\
$ -2.92 $ &    74   &   60&$ -0.62 $ &  3329   & 6361\\
$ -2.88 $ &   107   &   83&$ -0.57 $ &  3025   & 5780\\
$ -2.83 $ &   152   &  107&$ -0.52 $ &  2691   & 5142\\
$ -2.78 $ &   145   &  100&$ -0.47 $ &  2495   & 4728\\
$ -2.72 $ &   208   &  142&$ -0.43 $ &  2160   & 4006\\
$ -2.67 $ &   244   &  158&$ -0.38 $ &  1949   & 3516\\
$ -2.62 $ &   303   &  189&$ -0.32 $ &  1534   & 2768\\
$ -2.58 $ &   401   &  241&$ -0.28 $ &  1302   & 2349\\
$ -2.53 $ &   440   &  262&$ -0.22 $ &  1012   & 1826\\
$ -2.47 $ &   564   &  295&$ -0.17 $ &   728   & 1313\\
$ -2.42 $ &   656   &  296&$ -0.12 $ &   559   &  992\\
$ -2.38 $ &   822   &  364&$ -0.08 $ &   451   &  814\\
$ -2.33 $ &   988   &  395&$ -0.03 $ &   353   &  662\\
$ -2.28 $ &  1152   &  438&$ +0.03 $ &   287   &  551\\
$ -2.22 $ &  1300   &  489&$ +0.08 $ &   244   &  468\\
$ -2.17 $ &  1557   &  539&$ +0.12 $ &   185   &  355\\
\hline
\end{tabular}
\end{table}

\begin{figure}
\centering
\resizebox{75mm}{!}{\includegraphics{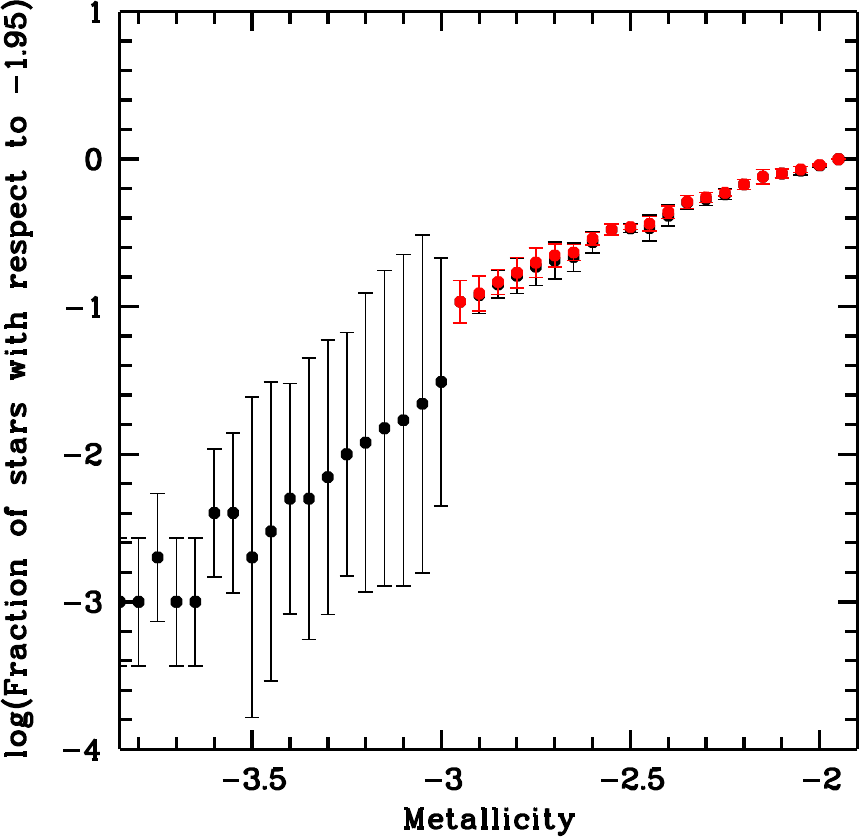}}
\caption{Average metal-weak tail of the MDF, normalised at metallicity
  $-1.95$: black present paper (corrected) and Hamburg/ESO
  (uncorrected) red; the H3 is also included in the
  average.\label{figavermdf}}
\end{figure}

\begin{table}
\caption{Average MDF, normalised at metallicity $-1.95$.  Aver2 is the
  average of the present MDF (corrected) and the Hamburg/ESO
  (uncorrected), $\sigma_2$ is the standard deviation, Aver3 is the
  average of the previous two and the H3 MDF, and $\sigma_3$ is the
  standard deviation.\label{tabavermdf}}
\begin{tabular}{ccccc}
\hline
[M/H] & Aver2 & $\sigma_2$ & Aver3 $\sigma_3$\\
\hline
 $-4.30$ & 0.000 &       &       &       \\
 $-4.25$ & 0.000 & 0.001 &       &      \\
 $-4.20$ & 0.000 & 0.001 &       &      \\
 $-4.15$ & 0.000 & 0.001 &       &      \\
 $-4.10$ & 0.001 & 0.001 &       &      \\
 $-4.05$ & 0.001 & 0.001 &       &      \\
 $-4.00$ & 0.000 & 0.001 &       &      \\
 $-3.95$ & 0.000 & 0.001 &       &      \\
 $-3.90$ & 0.001 & 0.001 &       &      \\
 $-3.85$ & 0.001 & 0.001 &       &      \\
 $-3.80$ & 0.001 & 0.001 &       &      \\
 $-3.75$ & 0.002 & 0.002 &       &      \\
 $-3.70$ & 0.001 & 0.001 &       &      \\
 $-3.65$ & 0.001 & 0.001 &       &      \\
 $-3.60$ & 0.004 & 0.004 &       &      \\
 $-3.55$ & 0.004 & 0.005 &       &      \\
 $-3.50$ & 0.002 & 0.005 &       &      \\
 $-3.45$ & 0.003 & 0.007 &       &      \\
 $-3.40$ & 0.005 & 0.009 &       &      \\
 $-3.35$ & 0.005 & 0.011 &       &      \\
 $-3.30$ & 0.007 & 0.015 &       &      \\
 $-3.25$ & 0.010 & 0.019 &       &      \\
 $-3.20$ & 0.012 & 0.028 &       &      \\
 $-3.15$ & 0.015 & 0.037 &       &      \\
 $-3.10$ & 0.017 & 0.044 &       &      \\
 $-3.05$ & 0.022 & 0.058 &       &      \\
 $-3.00$ & 0.031 & 0.060 &       &      \\
 $-2.95$ & 0.108 & 0.036 & 0.108 & 0.036\\
 $-2.90$ & 0.120 & 0.035 & 0.123 & 0.034\\
 $-2.85$ & 0.142 & 0.031 & 0.147 & 0.029\\
 $-2.80$ & 0.162 & 0.045 & 0.170 & 0.040\\
 $-2.75$ & 0.187 & 0.055 & 0.199 & 0.046\\
 $-2.70$ & 0.206 & 0.059 & 0.223 & 0.040\\
 $-2.65$ & 0.217 & 0.049 & 0.233 & 0.029\\
 $-2.60$ & 0.274 & 0.046 & 0.289 & 0.029\\
 $-2.55$ & 0.333 & 0.030 & 0.334 & 0.030\\
 $-2.50$ & 0.340 & 0.022 & 0.347 & 0.013\\
 $-2.45$ & 0.342 & 0.068 & 0.365 & 0.041\\
 $-2.40$ & 0.415 & 0.069 & 0.437 & 0.042\\
 $-2.35$ & 0.509 & 0.054 & 0.514 & 0.052\\
 $-2.30$ & 0.537 & 0.053 & 0.551 & 0.042\\
 $-2.25$ & 0.577 & 0.048 & 0.592 & 0.031\\
 $-2.20$ & 0.673 & 0.054 & 0.675 & 0.053\\
 $-2.15$ & 0.761 & 0.083 & 0.761 & 0.083\\
 $-2.10$ & 0.799 & 0.060 & 0.800 & 0.060\\
 $-2.05$ & 0.835 & 0.057 & 0.852 & 0.039\\
 $-2.00$ & 0.906 & 0.025 & 0.914 & 0.016\\
 $-1.95$ & 1.000 & 0.000 & 1.000 & 0.000\\
\hline
\end{tabular}
\end{table}

Above metallicity $-3.0$ the H3, the {\em uncorrected} HES, and our
{\em corrected} MDF are essentially identical. If we make the
assumption that our bias-corrected MDF is ground truth, then the
metallicity bias of the H3 and HES MDFs must be small, and in fact
negligible. We thus conclude that it is legitimate to use the average
of these three MDFs for comparison with theoretical models.{
The assumption we are making is equivalent to stating
that the bias corrections provided by \citet{Schoerck} and
\citet{Youakim} are not adequate to describe the actual
bias in their stellar sample. 
The bias corrections proposed by \citet{Schoerck} are based on models, 
while those of \citet{Youakim} are based on the statistical
properties of the sample. 
One advantage of using the average is that in each metallicity bin the
standard deviation can be used as an error estimate on the MDF. 
A further advantage is that the averaging process  makes the statistical errors
smaller.} The
H3 MDF is essentially undefined below $-3.0$; therefore, we suggest only
using the average of our MDF and that of HES in this metallicity
range. In Fig.\,\ref{figavermdf}, we show the two average values with
the standard deviation in each bin. The values are provided in Table
\ref{tabavermdf}.

\section{3D bias and metallicity maps\label{metmaps}}

We can now take advantage of the precise measurements of metallicity
and position for our large stellar sample to study the spatial
distribution of stars with different metallicities as has been
suggested by cosmological models \citep[e.g.][]{salvadori10}.  To
achieve this goal, we need to derive the 3D bias function, meaning that we also
need to study how the bias function depends on the position of the
stars. To this aim, we adopt the cylindrical coordinates $|z|$ and
$\rho = \sqrt{\left(R^2_{GC}-z^2\right)}$ and divide the space in
annuli of radial width $dR = 0.5$~kpc and height $dz=0.3$ kpc. In each
annulus, the photometric MDFs from the spectroscopic and photometric
samples have been derived and used to compute the local bias
function by adopting metallicity bins of 0.25\,dex.  In
the following, we use $Fe$ to denote the metallicity, as defined
in Sect.\,\ref{sec:sel}. 
From a global point of view, this means that we have constructed a 3D
bias function, $\mathrm{Bias} = \mathrm{Bias}(Fe, \rho, |z|)$,
that we can use to compute the correct local MDF and to construct
stellar metallicity maps.

\begin{figure*}
\centering
\resizebox{18.5cm}{!}{\includegraphics[clip=true]{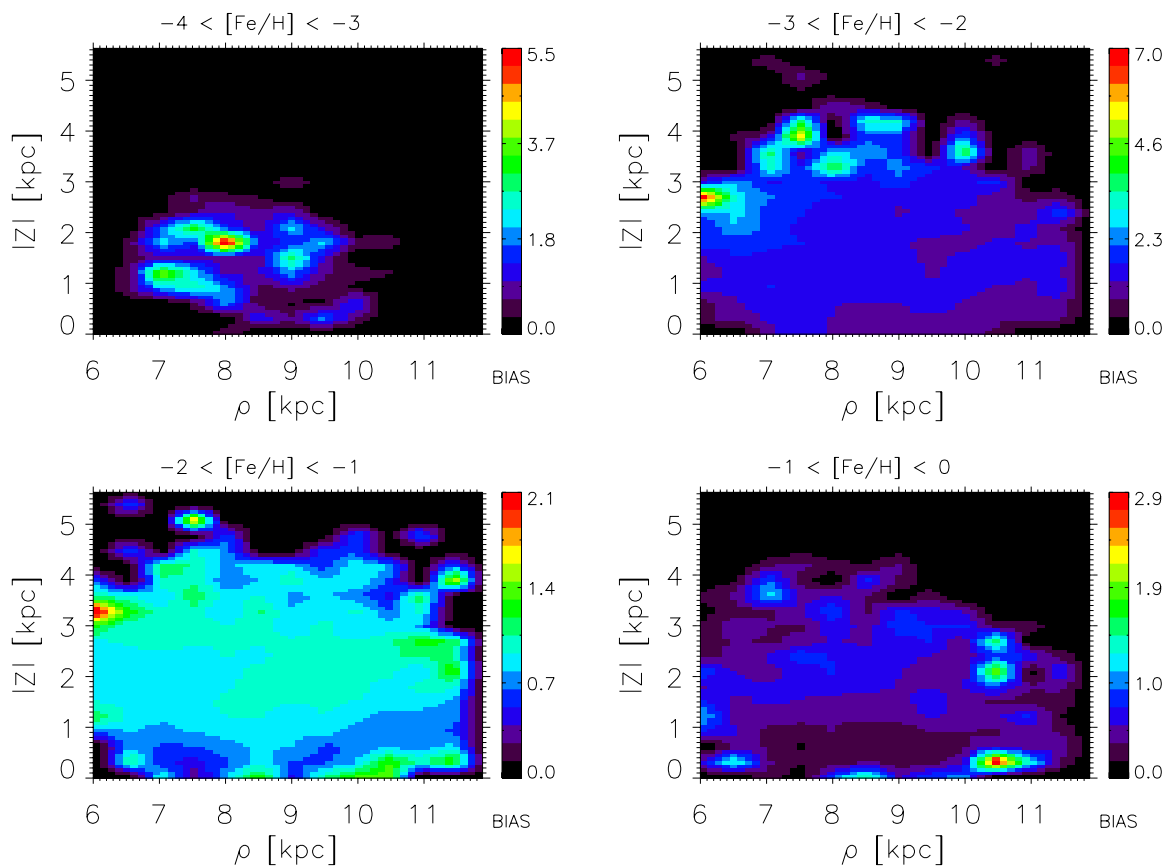}}
\caption{Bias maps in the cylindrical coordinate plane
  ($\rho$,$|z|$). Each panel shows the results for different stellar
  populations (see labels). \label{bias_map}}
\end{figure*}

In Fig.~\ref{bias_map}, we show a simplified view of the 3D bias
function obtained by dividing the stellar populations into different
metallicity ranges: extremely metal-poor stars ($-4\leq\mathrm{[Fe/H]}< -3$,
upper left panel), very metal-poor stars ($-3\leq\mathrm{[Fe/H]}< -2$, upper
right), metal-poor stars ($-2\leq\mathrm{[Fe/H]}< -1$, lower left), and
metal-rich stars ($-1\leq\mathrm{[Fe/H]}< 0$, lower right). Colours in the maps
illustrate the value of the bias obtained in each annulus, which is
identified by a single pixel in the ($\rho, |z|$) space. A bias equal
to zero means that there are no stars from the spectroscopic sample
within the pixel, which is indeed the case for extremely metal-poor
stars (upper left) located in the outermost regions ($|z|>3$~kpc,
$\rho>10$\,kpc).

Bearing in mind that the colour-bar scale is different in each panel,
we can notice several features in Fig.~\ref{bias_map}. First, the
contrast between the minimum and maximum values of the bias is higher
for very and extremely metal-poor stars ($\mathrm{[Fe/H]}< -2$, upper
panels), where it reaches values of $\mathrm{Bias} > 5$ in several
small and isolated regions. Our analysis shows that these extreme bias
values are a result of the small statistic in the considered
pixel. Conversely, the bias is much more uniform for metal-poor stars
(lower left panel), only spanning an interval of two\ in spite of the
more extended regions covered by the map. The contrast between minimum
and maximum bias increases again for metal-rich stars, as seen in the
lower right panel of Fig.~\ref{bias_map}, where we can notice three
zones with $\mathrm{Bias}\geq 2$. The high contrast between the bias
in these isolate small regions, and the rest of the space is at the
origin of the oscillations seen in the (1D) bias function analysed in
Sect. 5.3 (Fig.~\ref{bias}).

\begin{figure*}
\centering
\resizebox{18.5cm}{!}{\includegraphics[clip=true]{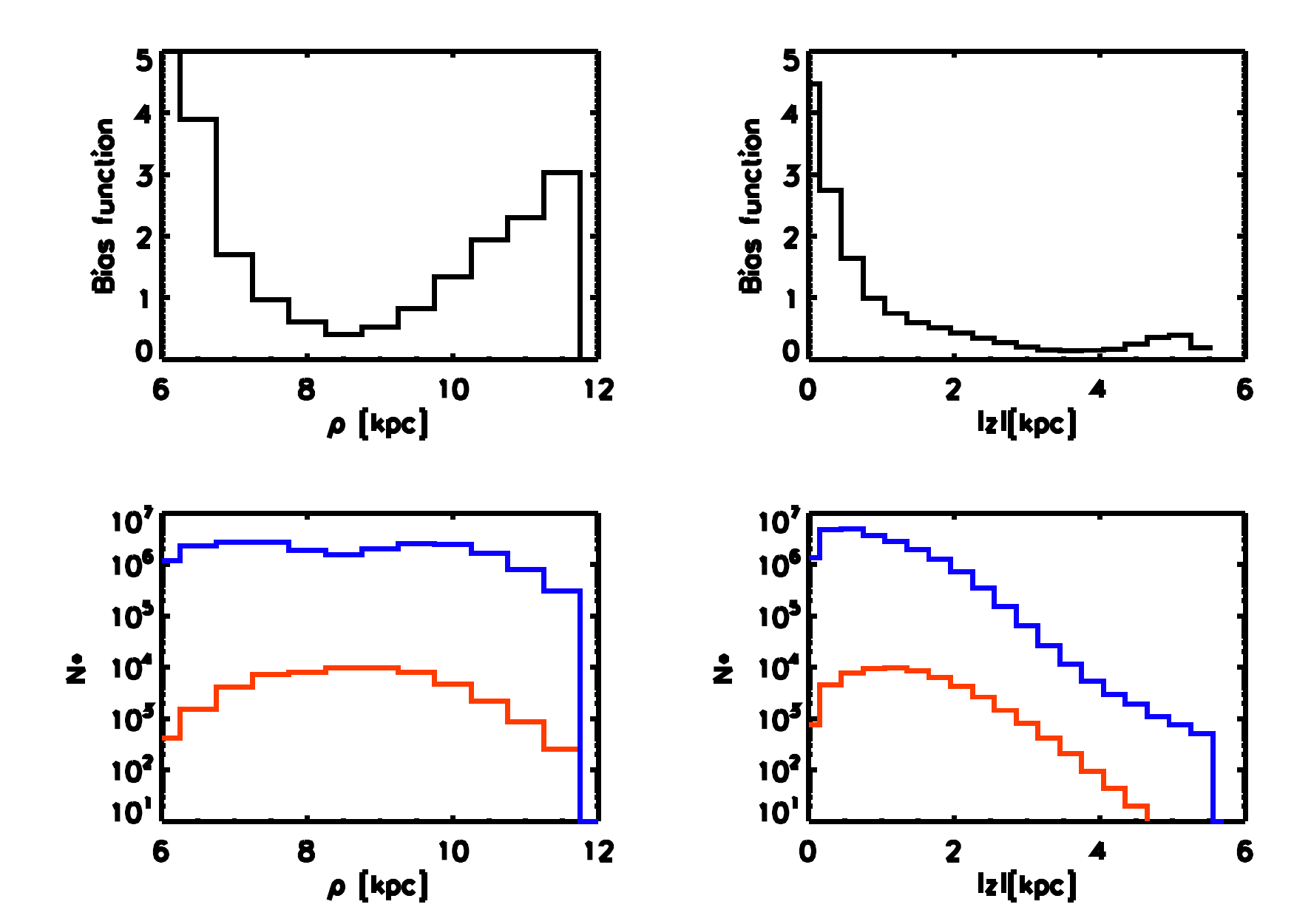}}
\caption{Variation of the bias function with the cylindrical radius,
  $\rho$ (left upper panel), and the height above the Galactic plane,
  $|z|$ (right upper panel). The lower panels show the distribution of
  stars from the photometric (blue) and spectroscopic (red) samples as
  functions of $\rho$ (left) and $|z|$ (right). \label{1d_biases}}
\end{figure*}

A more intuitive view of the bias function showing its dependency on
the single variables $\rho$ and $|z|$ is presented in
Fig.~\ref{1d_biases} (upper panels), while we refer the reader to
Fig.~\ref{bias} (Sec. 5.3) for the metallicity dependence. The 1D
bias functions shown in Fig.~\ref{1d_biases} were obtained
starting from the spectroscopic and photometric samples from which we
derived the radial and height stellar distributions displayed in the
same figure (lower panels). The upper left panel of
Fig.~\ref{1d_biases} shows how the bias function varies with the
radius $\rho$. We notice that $\mathrm{Bias}(\rho)$ is maximum at the lowest
radii, and it rapidly declines towards larger $\rho$, reaching its minimum
at $\rho\approx 8.5$~kpc, and then it rises again. This behaviour can
be easily understood: the position of the minimum corresponds to the
position of the Sun, $\rho=8.43$~kpc (see Sect.\,\ref{sec:comp_phot}),
and hence it identifies the radius where the number of stars in the
spectroscopic sample is larger, $N*\approx 10^4$, as can be seen in
the lower left panel of Fig.~\ref{1d_biases}. In the same panel, we
see that both at smaller and larger radii the number of stars in the
spectroscopic sample decreases, reaching the value $N*\leq 400$ at the
extremes. Conversely, the number of stars in the photometric sample
remains roughly at the constant value of $N* > 10^6$ for $6.5$~kpc $<
\rho\leq10$~kpc and slowly declines at larger $\rho$, following the
same trend of the spectroscopic sample. This behaviour drives the
shape of the bias function, which we remind the reader is obtained by dividing
the photometric and the spectroscopic metallicity distributions (see
Sect.~5.2 for more details).

The variation of the bias function with $|z|$ is shown in the right
upper panel of Fig.~\ref{1d_biases}. We see that $\mathrm{Bias}(|z|)$ is
maximum on the Galactic plane and then it rapidly decreases at higher
$|z|$, reaching values close to zero for $|z|\approx 4$~kpc. We can
also notice a little bump at $|z|\approx 5$~kpc. Once again, these
trends can be understood by looking at the effective number of stars
from the spectroscopic and photometric samples in each $|z|$ bin
(lower right panel of Fig.~\ref{1d_biases}). We see that the number of
stars in the spectroscopic sample gradually increases with $|z|$,
reaching the maximum $N*\approx 10^4$,at $|z|\approx 1$~kpc, and then
rapidly declining at higher $z$. Conversely, the number of stars in
the photometric sample is maximum close to the Galactic plane,
$|z|\leq0.5$~kpc, and then it starts to decline. This explains the trend
of the bias function with $|z|$.

We note that at $|z| > 4$~kpc the spectroscopic sample is very poorly
represented, $N*< 100$, which means that the bump observed in
$\mathrm{Bias}(z)$ at $|z|\approx 5$~kpc is driven by a very small
number of stars, $N*< 10$. To understand the origin of this bump we
should inspect the complete 3D bias maps shown in
Fig.~\ref{bias_map} again.  With the exception of{  metal-poor stars}
(left lower panel), all stellar populations show zero bias at $|z|\geq
4$~kpc for all $\rho$. This means that there are no stars in
the spectroscopic sample in these regions and at these metallicities.
At $|z|\approx 5$~kpc ($\rho\approx 7.5$~kpc) instead, very metal-poor
stars show a small isolated `island' with $\mathrm{Bias}\approx
1.5$. It is likely this small region, counting $<10$ stars, which
originates the small bump, observed $\mathrm{Bias}(z)$ at $|z|=5$
(right upper panel).

\begin{figure*}
\centering
\resizebox{18.5cm}{!}{\includegraphics[clip=true]{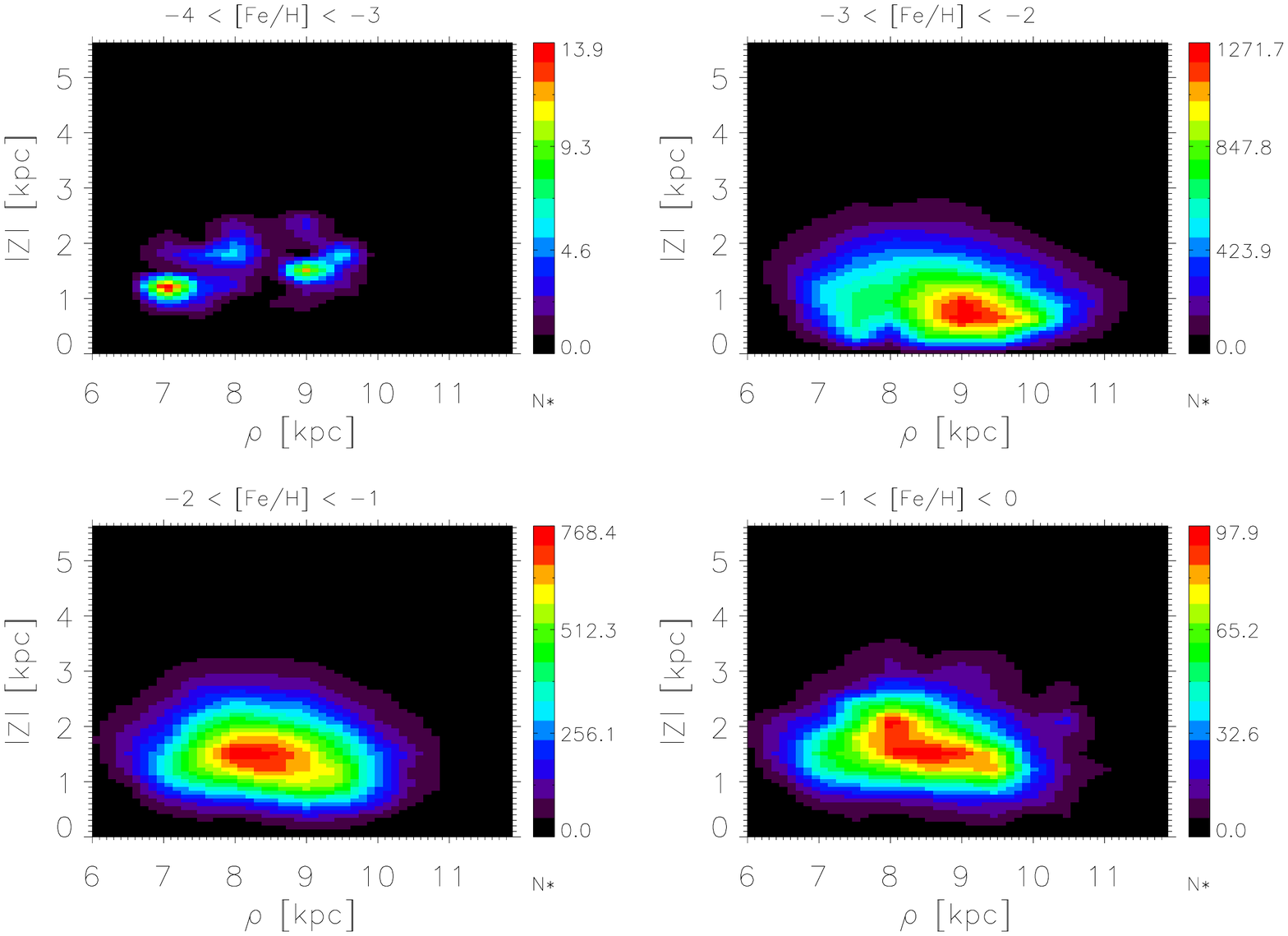}}
\caption{Distribution of extremely metal-poor (upper left), very
  meta-poor (upper right), metal-poor (lower left), and metal-rich
  (lower right) stars in the cylindrical coordinate plane
  ($\rho$,$|z|$). The distributions have been corrected using the 3D
  bias function.
\label{stars_map}}
\end{figure*}

In Fig.~\ref{stars_map}, we show the spatial distribution of extremely
metal-poor, very metal-poor, metal-poor, and metal-rich stars
corrected for the bias. In other words, these maps were obtained
by multiplying the raw spatial distributions of the spectroscopic
sample for the local 3D bias function. In Fig.~\ref{stars_map}, we can
first notice that the spatial distribution of extremely metal-poor
stars (EMPs, left upper panel) is completely different with respect to
those of the more metal-rich stellar populations (i.e.
$\mathrm{[Fe/H]}> -3$). EMPs are clustered in a smaller region of the
($\rho, |z|$) plane, and their spatial distribution does not appear
uniform like the others do. This is most likely a consequence of the
low number of statistics: in the three main regions filled by EMPs, we can
indeed see that the number of stars per pixel is $<15$.  Conversely,
more metal-rich stars cover almost the same regions of the ($\rho,
|z|$), and they are all much better represented in each pixel, where
$N*>30$ most of the time. However, we can notice that the maxima
of the stellar distributions (red regions) reside in three different
locations, which implies that there are spatial gradients in the
metallicity distribution function. A more in-depth and quantitative
analysis of these results, which includes a comparison with
cosmological models, is beyond the scope of the present paper and
will be presented in our future work (Salvadori et al., in prep).

\section{Dynamical properties of the SDSS spectroscopic sample\label{dynamics}}

In order to discuss the dynamical properties of the sample, we only make use of the good parallax sample.  In terms of Galactic location, the
two samples occupy approximately the same space in the $R_{GC}$,$z$
plane, as shown in Fig.\,\ref{rgc_z}. We corrected the Gaia parallaxes
for the global zero point of $-0.029$\,mas \citep{lindegren18}. With
these parallaxes, the Gaia proper motions, and the SDSS radial
velocities, we used the {\tt galpy}
code\footnote{\href{http://github.com/jobovy/galpy}{http://github.com/jobovy/galpy}}
\citep[][]{galpy} to evaluate orbital parameters and actions under the
Staeckel approximation \citep[][]{binney12,mackereth18} and using the
{\tt MWPotential2014} potential \citep[][]{galpy}.

\subsection{The Gaia-Sausage-Enceladus structure}

From the studies of \citet{belokurov18},
\citet{haywood18},\citet{helmi18} and \citet{dimatteo19},
we know that the Galactic halo\footnote{We loosely define the Galactic
  halo as the stars with a total space velocity higher than 180\kms
  \citep{dimatteo19}.}  is dominated by stars that were accreted from
a galaxy that we refer to as Gaia-Sausage-Enceladus (GSE).  The
precise dynamical definition of this object is still not known.  We
refrain from identifying, in dynamical spaces, multiple accretion
events \citep[][]{Naidu20} and small sub-structures, since we are aware
that, from the theoretical point of view, a single massive accretion
event can give rise to several dynamical sub-structures
\citep{Jean-Baptiste}.  We therefore concentrate on
isolating GSE, although we are aware that it is practically impossible
to identify a `pure' sample of GSE stars, since one of the outcomes
of the accretion is that stars that come from both the accreted
satellite galaxy and the accreting galaxy will end up in the same
region of any dynamical space.  Chemical information may be helpful to
disentangle the accreted from the accreting, but our information is
limited to an average metallicity and is not sufficient for this
purpose.

\begin{figure}
\centering
\resizebox{75mm}{!}{\includegraphics[clip=true]{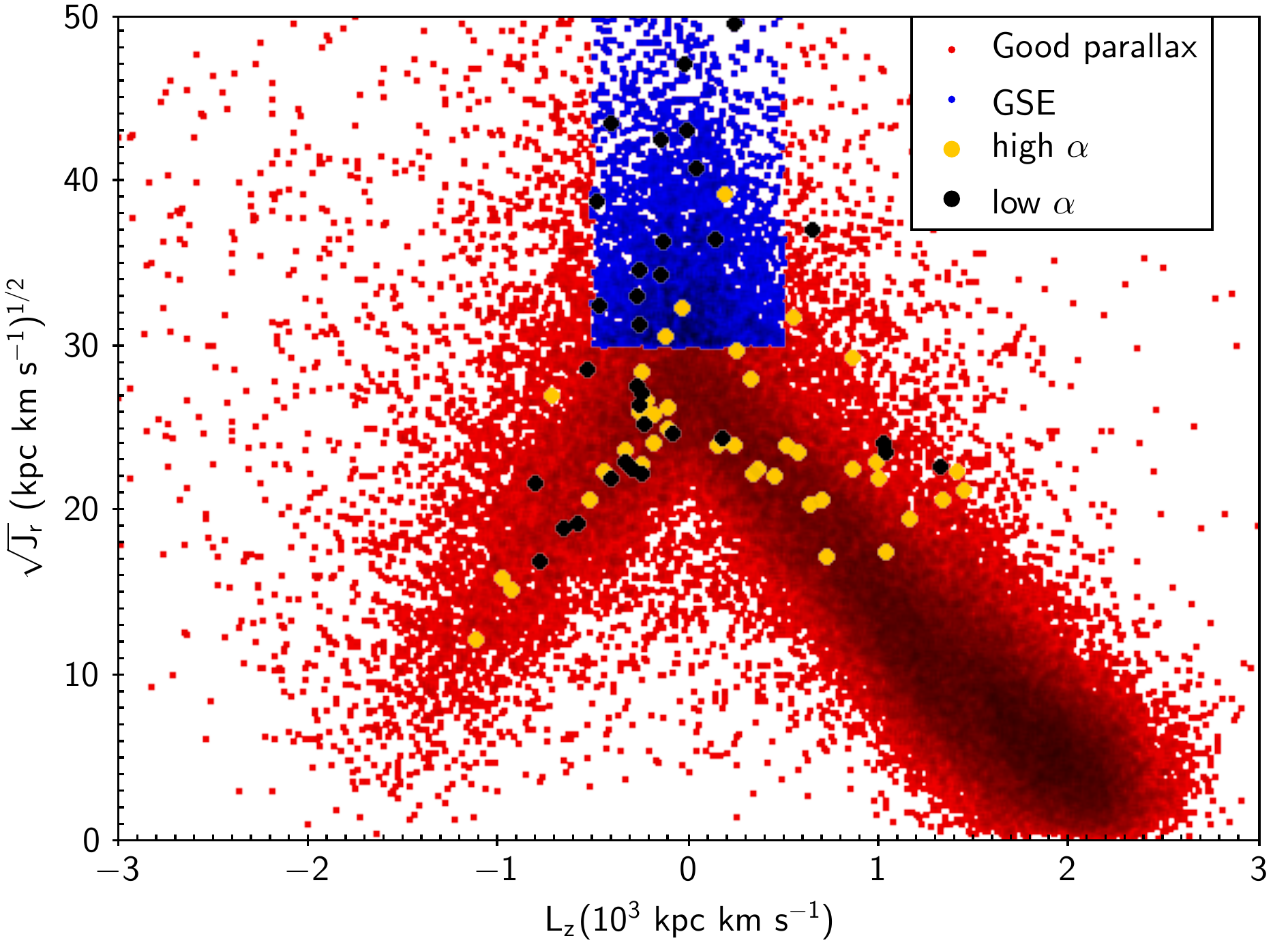}}
\caption{Diagram of angular momenta and square root of the radial
  action. Our good parallax sample is shown via red dots, the blue
  rectangle is the GSE selection proposed by \citet{Feuillet}, the
  yellow dots are the high-$\alpha$ stars of \citet{NS10}, while the
  black dots are their low-$\alpha$ stars.\label{LzJr}}
\end{figure}

Based on dynamical data and photometric
metallicities, \citet{Feuillet} suggested that the best way to select a pure GSE
sample is in the space of angular momentum $L_z$ and radial action
($J_r$, or its square root). In Fig.\,\ref{LzJr}, we show such a
diagram for our good parallax sample. The points in blue correspond to
the GSE selection of \citet{Feuillet}, namely $-500 \,{\rm
  kpc\ km\ s^{-1}} < L_z < 500\, {\rm kpc\ km\ s^{-1}} $ and $30\,{\rm
  (kpc\ km\ s^{-1})^{1/2}} < \sqrt{J_r} < 50\,{\rm
  (kpc\ km\ s^{-1})^{1/2}}$.

The existence of two sequences of [$\alpha$/Fe] in the solar
neighbourhood as a function of metallicity was initially discovered by
\citet[][and subsequent papers]{Fuhrmann98} and later confirmed by
\citet{NS10}, who introduced the terms low-$\alpha$ and high-$\alpha$
stars.  \citet{haywood18} demonstrated convincingly that the
low-$\alpha$ stars should be associated with GSE, while high-$\alpha$
are stars formed in the Milky Way. This was also later confirmed by
\citet{helmi18}. In Fig.\,\ref{LzJr}, we also plot the stars
classified as low $\alpha$ and high $\alpha$ by \citet{NS10}.
Seventeen of the \citet{NS10} stars fall in the \citet{Feuillet} GSE
selection box; of these, fourteen are low-$\alpha$ stars and three are
high-$\alpha$ stars.  This allows us to estimate that the
\citet{Feuillet} GSE selection box includes about 18\,\% of stars that
are not associated with GSE. This confirms the theoretical expectation
that it is impossible to dynamically select a pure sample of the
accreted stars.  In the case of the \cite{NS10} stars, the chemical
information allows one to disentangle the two populations; however, the
two $\alpha$/Fe sequences merge at metallicities lower than about
$-1.0$, making this chemical diagnostic unusable. Pragmatically, we
decided to adopt the \citet{Feuillet} selection, keeping in mind that
there will be a contamination, since this appears to be the least
contaminated selection of GSE.  We would also like to point out that
there is also no guarantee that this selection of GSE provides an
unbiased sample of the stars of the accreted galaxy.

The \citet{Feuillet} selection contains 3\,227 stars in our good
parallax sample. 
Since this sample was extracted from a biased sample of stars,
this MDF cannot be taken as representative of the MDF of GSE.  On the
other hand, we cannot apply the bias correction that has been derived
for the global sample, since this sub-sample has, a priori, a
different bias. We cannot apply the \citet{Feuillet} selection to the
photometric sample to derive an appropriate bias function, since we
lack the radial velocities necessary to derive $L_z$ and $J_r$.
However, in Sect.\,\ref{metmaps} we derive the space dependence
of the bias function. We can therefore divide the GSE sample into bins
in the $\rho,|z|$ space and apply the appropriate bias function to
each bin.

\begin{figure}
\resizebox{75mm}{!}{\includegraphics{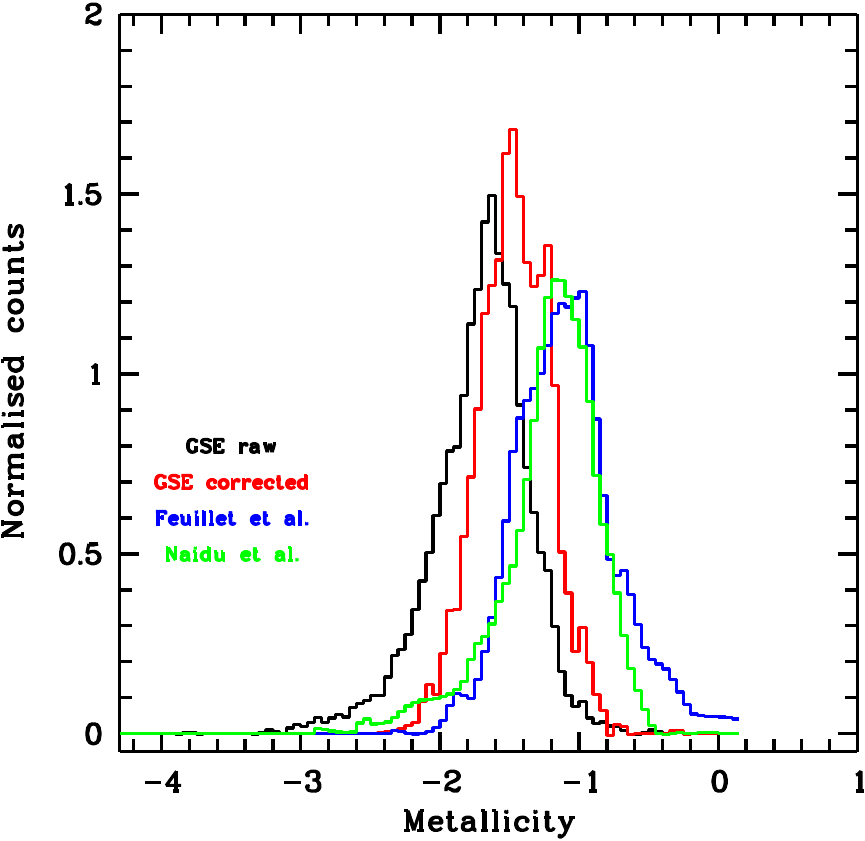}}
\caption{Observed metallicity histogram of GSE, for the whole sample
  (black), and the bias-corrected histogram (red). The histogram of
  \citet{Feuillet} is shown in blue, while that of
  \citet{Naidu20} is shown in green.\label{histogse_corr}}
\end{figure}

The bias correction results in shifting the MDF to higher
metallicities, as expected; however, our MDF remains distinctly more
metal-poor than that derived by \citet{Feuillet}, as shown in
Fig. \ref{histogse_corr}.  Both our raw and bias-corrected MDF{ of GSE}
lack the
significant population between $-1.0$ and solar metallicity that is
present in the \citet{Feuillet} MDF{ of GSE}. 
Our MDF is also slightly narrower
than that of \citet{Feuillet}. Although the shape of none of the MDFs
is Gaussian, a Gaussian fit can provide a measure of their width. Our
MDFs (both raw and bias-corrected) have a 
full width at half maximum of about 0.6\,dex, while
that of \citet{Feuillet} is about 0.7\,dex.

A possible reason for the discrepancy is that the \citet{Casagrande19}
photometric calibration, which is used by \citet{Feuillet} to derive
the metallicities, is validated only down to $-2.0$. Thus, the
\citet{Feuillet} sample may have been selected with a bias against
metal-poor stars.  To gain some insight into a comparison of our
metallicities and those obtained from the \citet{Casagrande19}
calibration, we cross-matched our sample with Skymapper DR~1.1
\citep{skymapper1} and with 2MASS \citep{2MASS}. We found a
sub-sample of 20\,203 unique stars, and we applied the \citet[][Eq.
  12]{Casagrande19} calibration. This photometric metallicity
correlates only weakly with our spectroscopy metallicity. The
metallicity distributions of the same stars using the two different
metallicity indicators do not resemble each other. The photometric
metallicity is on average 0.7\,dex more metal rich, but with a
dispersion that is of the same order of magnitude.  We do not want to
push this comparison further, since our stars are dwarfs, while those
used by \citet{Feuillet} are giants, and the systematics may well be
different.

\citet{Naidu20} also provide an MDF for GSE, and this is shown in
Fig.\,\ref{histogse_corr}. Their MDF is similar to that of
\citet{Feuillet} around the peak, but it departs from 
it both in the metal-weak and in the 
metal-rich tail. It is clearly different from ours, that has many more
metal-poor stars.  We point out that the \citet{Naidu20} selection of
GSE is very different from that of \citet{Feuillet} and adopted by us.
In fact, they write the following: {\em \emph{``We select GSE stars by excluding previously
  defined structures and requiring} $e> 0.7$ ''}. Hence, it is necessary to
have previously isolated all the different sub-structures identified by
\citet{Naidu20}. In order to mimic the \citet{Naidu20} selection, we selected stars with $e > 0.7$ and $-500 \,{\rm
  kpc\ km\ s^{-1}} < L_z < 500\, {\rm kpc\ km\ s^{-1}} $ in our sample. This
translates, in our preferred selection plane, to $18\,{\rm
  (kpc\ km\ s^{-1})^{1/2}} < \sqrt{J_r} < 60\,{\rm
  (kpc\ km\ s^{-1})^{1/2}}$. As stated above, this increases the
contamination of stars not related to GSE, probably above the level of
50\%, as estimated from applying this selection to the \citet{NS10}
sample. It is, however, not clear if these contaminating stars have an
MDF different from GSE or not.  We applied this selection to our
sample, and the MDF we obtain is similar to what we obtain with the
\citet{Feuillet} selection.  As in the case of \citet{Feuillet}, we may
suspect that the metallicity scale of \citet{Naidu20} is different
from ours. This seems unlikely since the metal-weak tail of our
bias-corrected MDF agrees very well with that of \citet{Naidu20}. A
more detailed comparison will have to wait until the H3 catalogue is
fully published and metallicities of known stars can be compared.

We can, however, comment on the similarity of the 
peak position between \citet{Feuillet} and \citet{Naidu20},
in spite of their disagreement in both metal-weak and
metal-rich tails.
Both \citet{Feuillet} and \citet{Naidu20}
rely on giant stars. However, neither takes into account
the fact that the rate at which the stars leave the
main sequence is metallicity dependent (the {\em \emph{evolutionary flux}};
see \citealt{RB86}) or the fact that the rate of evolution
along the red giant branch is also metallicity dependent
\citep{Rood72}. These effects were discussed in the case
of the Galactic bulge by \citet{Zoccali}.
It may well be that the similarity in the peak of the MDFs of GSE
of the two above-mentioned studies does in fact signal
that both these MDFs are biased by the choice of tracers, namely
giant stars.\footnote{
After the acceptance of the paper, C. Conroy brought to our attention
the fact that Naidu et al. (2000) accounted for the evolutionary flux
via their selection function. 
However, the MDFs shown in their figures, from which we have taken
the values shown in our figures, are all `raw'.
C. Conroy (priv. comm.)
evaluated the metallicity dependence of an MDF based
on giants, using the mock Gaia catalogue of \citet{Gaiamock} 
and estimates the effect to be of the order of about 0.05\,dex, which is
far too small to account for the difference between our MDF of GSE
and that of \citet{Naidu20}.    
}

\begin{figure}
\resizebox{75mm}{!}{\includegraphics{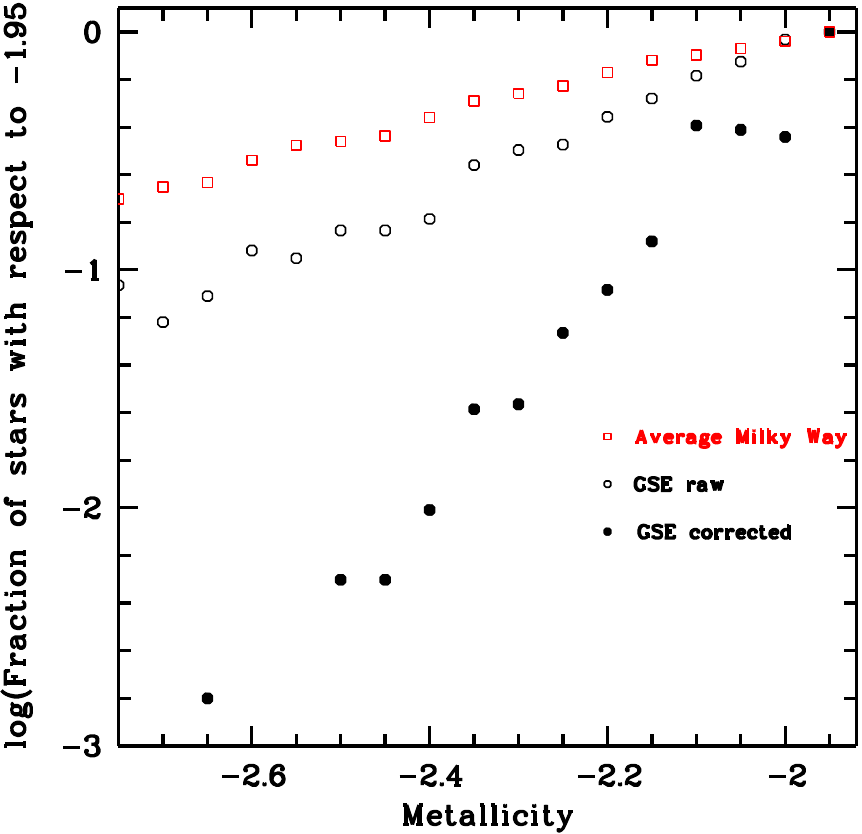}}
\caption{Observed metallicity histogram of GSE, normalised 
to $-1.95$ compared to that of the global sample.{ The global sample is bias corrected as described in 
Sect.\,\ref{sec:bias}. The GSE sample is shown both raw
and corrected, as described in Sect.\,\ref{metmaps}.}
\label{histogse_n2}}
\end{figure}

It is remarkable that the slope of the metal-weak tail of the MDF of
GSE is much steeper than what was obtained from the full sample, as
shown in Fig.\,\ref{histogse_n2}.  This is true even if we consider
our observed MDF, before bias correction; thus, it must be considered
as a very robust finding.  It could be argued that rather than
comparing the GSE MDF to our mean MDF, which also relies on our 1D-corrected MDF, we should have used our 3D-corrected MDF for the full
sample. There are two reasons for not doing this. First of all, in the
metallicity range that is relevant for GSE, the 1D-corrected and 3D-corrected MDFs are pretty close, while they become increasingly
different at lower metallicities where the 3D bias is plagued by a low
number of stars per bin. Secondly, the {\em \emph{average}} MDF has a much
smaller error than our 1D-corrected MDF and, {\em \emph{a fortiori,}} than
our 3D-corrected MDF.

\subsection{Errors on dynamical quantities and consequences on selections}

In order to evaluate the uncertainty in the derived kinematical
parameters, we used the {\tt pyia}
code\footnote{\href{https://github.com/adrn/pyia}{https://github.com/adrn/pyia}
  \citep{pyia18}} to generate samples of the six parameters
(coordinates, parallax, proper motions, and radial velocity) in input
to {\tt galpy}\ for each object. Samples are generated from a
multi-variate Gaussian distribution, where the errors on the parallax,
proper motions, and radial velocity; and the correlation coefficients
between parallax and proper motions are only taken into account in the
construction of the covariance matrix. One thousand realisations were
extracted for each object. The samples were then fed into {\tt galpy,}
and the errors were evaluated as the standard deviations of the obtained
orbital parameters, after unbound orbits were rejected.  For the error
estimate, we only accepted objects for which more than 950 out of 1000
realisations provided meaningful results, namely 54\,215 stars over
57\,386.  When we consider the GSE sub-sample, the number of stars
retained drops from 3\,227 to 2\,487. We consider that stars that have
too few meaningful realisations to have uncertain orbital parameters.
   
The question is, what is the effect of these errors on a sub-sample
selected on the basis of dynamical quantities?  We compared the
metallicity histograms for the full GSE sample with errors (2\,487
stars) and for a sub-sample selected with the
supplementary condition $ eL_z < 263.983\,{\rm kpc\ km\ s^{-1}}$ and $
eJ_r < 995.473\,{\rm kpc\ km\ s^{-1}}$, where $eL_z$ and $eJ_r$ are
the errors on $L_z$ and $J_r$ derived as described above, and the
condition is fixed to the median of the error distribution.  From
this comparison, it is clear that the shape of the MDF is by and large
the same if we consider the full GSE sample or only the sub-sample
that has smaller errors on the orbital parameters.  We thus conclude
that the shape of the MDF for our GSE selection is robustly
determined.
In the same way, we find that the conclusions are not different if we
consider the full sample of 3\,227 or the sub-sample of 2\,487 stars
with reliable orbital parameters, according to the above definition.

\section{Discussion}

Our recommended metal-weak tail of the MDF is provided in Table
\ref{tabavermdf}. We deduced that the raw Hamburg/ESO MDF is
essentially unbiased from its similarity with our own bias-corrected
MDF and the H3 MDF.  Why would the HES be unbiased, when
\citet{Schoerck} warned that it is biased in favour of the metal-poor
stars and provide bias corrections?  Our understanding is that {\em
  \emph{globally}} the HES is biased; however, the follow-up on metal-poor
candidates selected from the objective prism spectra has been unbiased
for all stars of metallicity below $-2.0$.
Quite likely, this was because of the low resolution of the prism
spectra, which made it difficult to discriminate between a $-2.0$ and
a $-3.0$ star. For this reason, the HES MDF, normalised at $-1.95$, is
effectively unbiased.

Another question is, why does the Pristine MDF show a larger fraction
of metal-poor stars at any metallicity compared to the other
three MDFs? This needs to be further investigated using the
Pristine data. However, we think this may be due to two factors:
\citet{Youakim} did not use Gaia-based distance estimates to
  `clean' their sample from white dwarfs, but relied on colour cuts.
  Our own experience, discussed in Sect.\,\ref{sec:gaia}, shows how the sample
  appeared much more clear, manageable, and consistent with isochrones,
  once we applied a distance cut.
\citet{Youakim} combined the Pristine filter with SDSS wide-band
  photometry to derive their metallicities.  In
  Fig.\,\ref{pl_gz_bprp}, we show how the SDSS photometry can be discrepant
  with Gaia photometry. It could be that the excess in metal-poor
  stars is partially due to a fraction of stars with poor SDSS
  photometry.

{ Another possible issue to be considered is the 
lower sensitivity to metallicity of any photometric
metallicity based on a narrow-band filter centred on
\ion{Ca}{ii} lines, for warm TO stars, with respect
to cool giants. This effect is clearly
illustrated in Fig.\,11 of \citet{pristineV}.
Finally, there is a degeneracy between  metallicity
and surface gravity
that is not accounted for in \citet{Youakim} and that could 
cause part of the discrepancy.
} 
It would certainly be useful to re-derive the Pristine MDF after
cleaning the sample for stars of too-high absolute magnitude and
uncertain wide-band photometry.

The MDF we derived for the GSE sub-structure is clearly at odds with
that of \citet{Feuillet} and \citet{Naidu20}. We already pointed out
how there could be an offset in the metallicity scales. We
would like to point out that our sample and the \citet{Feuillet}
sample do not overlap perfectly in the sampled galactocentric volume.
The \citet{Feuillet} sample preferentially contains stars with $|z| <
1$\,kpc, while our sample is composed mainly of stars with $|z| >
1$\,kpc.  Also, the \citet{Naidu20} sample covers a very different
region in the $R_{GC},|Z|$, plane. In particular, their sample pushes
to 50\,kpc, obviously thanks to the use of giant stars as tracers,
while we only used TO stars. If there were any gradient of the
metallicity of GSE with $z$ or $R_{GC,}$ one would expect the three
samples to provide different MDFs.

\section{Conclusions}

We used our own analysis of SDSS spectra in order to determine
the MDF of the metal-poor component of the Milky Way. The metal-weak
part of our MDF is in excellent agreement with that determined by
\citet{Allende14} from an independent analysis of SDSS spectra. This
fact, together with the external comparison of a small sub-set of stars
analysed at higher resolution and/or S/N gives us confidence in
our spectroscopic metallicity, for which we estimate a precision of
0.1\,dex and an accuracy of 0.35\,dex. We showed that our sample is
biased, in the sense that metal-poor stars are over-represented.

Through a comparison with a sample of stars, that we assume
to be unbiased with respect to metallicty, for which we
determined
photometric metallicites, we derived the bias function. 
This allowed us to provide the unbiased MDF that
is given in Table\,\ref{tabmdf}.  When considering the metal-weak tail
of the MDF, normalised at $-1.95$, we find that our bias-corrected
MDF is essentially identical to that of \citet{Naidu20} and that of
\citet{Schoerck}.  We therefore advocate that the best estimate is
obtained by averaging these three MDFs. This average MDF and its
standard deviation is provided in Table\,\ref{tabavermdf}, and we
recommend using this table for comparisons with theoretical
predictions.

In the process of estimating the bias in our spectroscopic sample, we
introduced a new photometric metallicity estimate based on SDSS $u-g$
and $g-z$ colours and on distance estimates based on Gaia
parallaxes. These estimates, along with effective temperatures and
surface gravities, for 24\,037\,008 unique TO stars is available
through CDS\footnote{cdsurl}.

Thanks to the Gaia distances and proper motions, combined with SDSS
radial velocities, we computed Galactic orbits and actions for a
sub-set of our spectroscopic sample, characterised by good parallaxes.
From this sample, we selected likely members of the GSE structure.  We
were able to derive the metallicity bias of this GSE sample, looking
at the position of these stars in the galaxy. In this way, we derived an
MDF for GSE, as shown in Figs. \ref{histogse_corr} and
\ref{histogse_n2}.  The most notable result is that the fraction of
metal-poor stars in GSE is clearly lower than that observed in the
Milky Way. This implies a very different formation and evolution
history of GSE with respect to the Milky Way.  Furthermore, although
the 'progenitor' of GSE might contain more metal-poor stars \citep[see
  e.g.][]{Jean-Baptiste} than what we find in the GSE structure, our
result seems to suggest that the metal-poor tail of the Galactic halo
is not made by massive dwarf galaxies like the progenitor of GSE, but
most likely by smaller (ultra-faint) dwarfs. Indeed, according to
$\Lambda$CDM models, ultra-faint dwarfs are expected to be the basic
building blocks of more massive systems and the 
formation environment of extremely metal-poor stars \citep[see e.g.][and
  references therein]{Salvadori15}.

\begin{acknowledgements}

This paper is dedicated to the late Roger Cayrel who has inspired the
TOPoS survey as well as all of our work on metal-poor stars. He passed
away on January 11$^{th}$ 2021, and he will be deeply missed.  We are
grateful to Bertrand Plez who computed all the metal-poor model
atmospheres and associated synthetic spectra used in this paper.  We
acknowledge useful discussions with David Aguado on the contents of
this paper.  We gratefully acknowledge support from the French
National Research Agency (ANR) funded projects ``Pristine''
(ANR-18-CE31-0017) and ``MOD4Gaia" (ANR-15- CE31-0007).
A.J.K.H., H.-G.L. and N.C. gratefully acknowledge funding by Deutsche
Forschungsgemeinschaft (DFG, German Research Foundation) -- Project-ID
138713538 -- SFB 881 (``The Milky Way System''), subprojects A03, A04,
A05, and A11. S.S. acknowledges support from the ERC Starting Grant
NEFERTITI H2020/808240 and the PRIN-MIUR2017, The quest for the first
stars, prot. n. 2017T4ARJ5.

This work has made use of data from the European Space Agency (ESA)
mission {\it Gaia}
(\href{https://www.cosmos.esa.int/gaia}{https://www.cosmos.esa.int/gaia}),
processed by the {\it Gaia} Data Processing and Analysis Consortium
(DPAC,
\href{https://www.cosmos.esa.int/web/gaia/dpac/consortium}{https://www.cosmos.esa.int/web/gaia/dpac/consortium}). Funding
for the DPAC has been provided by national institutions, in particular
the institutions participating in the {\it Gaia} Multilateral
Agreement.

Funding for the SDSS and SDSS-II has been provided by the Alfred
P. Sloan Foundation, the Participating Institutions, the National
Science Foundation, the U.S. Department of Energy, the National
Aeronautics and Space Administration, the Japanese Monbukagakusho, the
Max Planck Society, and the Higher Education Funding Council for
England. The SDSS Web Site is
\href{http://www.sdss.org/}{http://www.sdss.org/}.

The SDSS is managed by the Astrophysical Research Consortium for the
Participating Institutions. The Participating Institutions are the
American Museum of Natural History, Astrophysical Institute Potsdam,
University of Basel, University of Cambridge, Case Western Reserve
University, University of Chicago, Drexel University, Fermilab, the
Institute for Advanced Study, the Japan Participation Group, Johns
Hopkins University, the Joint Institute for Nuclear Astrophysics, the
Kavli Institute for Particle Astrophysics and Cosmology, the Korean
Scientist Group, the Chinese Academy of Sciences (LAMOST), Los Alamos
National Laboratory, the Max-Planck-Institute for Astronomy (MPIA),
the Max-Planck-Institute for Astrophysics (MPA), New Mexico State
University, Ohio State University, University of Pittsburgh,
University of Portsmouth, Princeton University, the United States
Naval Observatory, and the University of Washington.

Funding for SDSS-III has been provided by the Alfred P. Sloan
Foundation, the Participating Institutions, the National Science
Foundation, and the U.S. Department of Energy Office of Science. The
SDSS-III web site is http://www.sdss3.org/.

SDSS-III is managed by the Astrophysical Research Consortium for the
Participating Institutions of the SDSS-III Collaboration including the
University of Arizona, the Brazilian Participation Group, Brookhaven
National Laboratory, Carnegie Mellon University, University of
Florida, the French Participation Group, the German Participation
Group, Harvard University, the Instituto de Astrofisica de Canarias,
the Michigan State/Notre Dame/JINA Participation Group, Johns Hopkins
University, Lawrence Berkeley National Laboratory, Max Planck
Institute for Astrophysics, Max Planck Institute for Extraterrestrial
Physics, New Mexico State University, New York University, Ohio State
University, Pennsylvania State University, University of Portsmouth,
Princeton University, the Spanish Participation Group, University of
Tokyo, University of Utah, Vanderbilt University, University of
Virginia, University of Washington, and Yale University.
\end{acknowledgements}
 
\bibliographystyle{aa}

\appendix

\section{Details on the selection of the spectroscopic sample \label{appa}}

In January 2016, we downloaded all the spectra from SDSS DR12 (SDSS \& BOSS)
with the following  SQL command:

{
\tiny
\tt
\begin{tabular}{l}
SELECT\\
  S.plate,S.MJD,S.fiberid,S.ObjID, S.ra, S.dec, S.type, S.flags as fl,\\
  S.psfMag\_u as u, S.psfMag\_g as g, S.psfMag\_r as r , S.psfMag\_i as i, \\
  S.psfMag\_z as z,\\
  S.extinction\_u as extu, S.extinction\_g as extg, S.extinction\_r as extr , S.extinction\_i as exti, \\
  S.extinction\_z as extz, S.survey as survey, S.programname as prog, \\
sp.TEFFADOP as teff, sp.TEFFADOPUNC as teff\_err, sp.LOGGADOP as glog\\
,\\
  sp.LOGGADOPUNC as glog\_err, sp.FEHADOP as feh, sp.FEHADOPUNC as feh\_err, \\
  sp.SpecObjId into mydb.dr12 from specPhotoAll as S JOIN sppParams as sp  ON S.SpecObjID = sp.SpecObjID\\
WHERE\\
  (S.type = 6)                                                  AND\\ 
  (S.flags \& dbo.fPhotoFlags('BRIGHT') = 0)                     AND  \\
  (S.flags \& dbo.fPhotoFlags('SATURATED') = 0)                  AND  \\
  (S.flags \& dbo.fPhotoFlags('BLENDED') = 0)                    AND  \\
  (S.flags \& dbo.fPhotoFlags('NOTCHECKED') = 0)                 AND  \\
  (S.flags \& dbo.fPhotoFlags('DEBLENDED\_AS\_MOVING') = 0)        AND\\
  (S.psfMag\_g-S.extinction\_g-S.psfMag\_z+S.extinction\_z between 0.18 and 0.70) AND\\
  (S.psfMag\_u-S.extinction\_u-S.psfMag\_g+S.extinction\_g > 0.7 ).\\
\end{tabular}
}

Apart from the flags used to make sure the object is classified as a star,
is not too bright or saturated, and so on, this boils down
to the conditions $0.18 \le (g-z)_0\le 0.70$ and $(u-b)_0> 0.7.$
These correspond to the conditions of the TOPoS Survey, which was selected from
SDSS DR9 and processed with {\tt abbo}.
This selection resulted in 312\,009 spectra of 266\,442 unique stars.

\section{Comparison samples}
\onecolumn

\begin{longtable}{lccrcccc}
\caption{Metallicities of stars in the TOPoS sample, observed at high resolution, compared to those derived by \mygi\ on the SDSS spectra.
\label{hr}
}
\\
\hline
\hline
Star  &  RA & DEC &    Ref. & \teff &      [Fe/H]  &  \teff &  [M/H]\\
      & \multicolumn{2}{c}{2000}              &     & K     &       dex    &   K    &    dex \\
      &     &     &         & Lit.  &       Lit.   & adop.  & \mygi \\
\hline
\endfirsthead
\caption{continued.}\\
\hline
\hline
Star  &  RA & DEC &    Ref. & \teff &      [Fe/H]  &  \teff &  [M/H]\\
      & \multicolumn{2}{c}{2000}             &     & K     &       dex    &   K    &    dex \\
      &     &     &         & Lit.  &       Lit.   & adop.  & \mygi \\
\hline
\endhead
\hline
\endfoot
 SDSS J$002113-005005$ & 00:21:14 & $-00$:50:05& 2  &    6546 &   $-3.15$ &   6398 & $-2.81$\\
 SDSS J$002558-101509$ & 00:25:59 & $-10$:15:09&4,8 &    6408 &   $-3.08$ &   6424 & $-2.96$\\
 SDSS J$002749+140418$ & 00:27:49 & $+14$:04:18& 2  &    6125 &   $-3.37$ &   6071 & $-3.59$\\
 SDSS J$003507-005037$ & 00:35:08 & $-00$:50:38&4,8 &    5891 &   $-2.95$ &   5886 & $-2.83$\\
 SDSS J$003730+245750$ & 00:37:30 & $+24$:57:51& 9  &    6350 &   $-3.42$ &   6350 & $-3.11$\\
 SDSS J$003954-001856$ & 00:39:55 & $-00$:18:57&4,8 &    6382 &   $-2.94$ &   6396 & $-3.05$\\
 SDSS J$004029+160416$ & 00:40:29 & $+16$:04:16& 2  &    6422 &   $-3.27$ &   6388 & $-2.87$\\
 SDSS J$012032-100106$ & 01:20:33 & $-10$:01:07&4,8 &    5804 &   $-3.50$ &   5798 & $-3.72$\\
 SDSS J$012125-030943$ & 01:21:25 & $-03$:09:44&4,8 &    6127 &   $-2.91$ &   6128 & $-3.46$\\
 SDSS J$012442-002806$ & 01:24:42 & $-00$:28:07&4,8 &    6273 &   $-2.79$ &   6281 & $-2.00$\\
 SDSS J$014036+234458$ & 01:40:36 & $+23$:44:58& 5  &    5848 &   $-4.00$ &   5884 & $-3.33$\\
 SDSS J$014721+021819$ & 01:47:22 & $+02$:18:20&4,8 &    5967 &   $-3.30$ &   5964 & $-3.82$\\
 SDSS J$014828+150221$ & 01:48:29 & $+15$:02:22&4,8 &    6151 &   $-3.41$ &   6154 & $-1.79$\\
 SDSS J$021238+013758$ & 02:12:38 & $+01$:37:58& 5  &    6333 &   $-3.59$ &   6300 & $-3.17$\\
 SDSS J$021554+063901$ & 02:15:54 & $+06$:39:01&4,8 &    6005 &   $-2.75$ &   6002 & $-3.41$\\
 SDSS J$031745+002304$ & 03:17:46 & $+00$:23:04& 2  &    5786 &   $-3.46$ &   5710 & $-3.27$\\
 SDSS J$035925-063416$ & 03:59:26 & $-06$:34:16&4,8 &    6281 &   $-3.17$ &   6290 & $-3.30$\\
 SDSS J$040114-051259$ & 04:01:15 & $-05$:12:59&4,8 &    5500 &   $-3.62$ &   5925 & $-3.14$\\
 SDSS J$074748+264543$ & 07:47:49 & $+26$:45:44&4,8 &    6434 &   $-3.36$ &   6452 & $-3.03$\\
 SDSS J$075338+190855$ & 07:53:39 & $+19$:08:56&4,8 &    6439 &   $-2.45$ &   6458 & $-2.40$\\
 SDSS J$080336+053430$ & 08:03:37 & $+05$:34:31&4,8 &    6360 &   $-2.94$ &   6373 & $-2.37$\\
 SDSS J$082118+181931$ & 08:21:18 & $+18$:19:32& 2  &    6158 &   $-3.80$ &   6161 & $-3.28$\\
 SDSS J$082452+163356$ & 08:25:11 & $+16$:35:00& 1  &    5463 &   $-3.22$ &   5946 & $-0.69$\\ 
 SDSS J$082521+040334$ & 08:25:21 & $+04$:03:34& 2  &    6340 &   $-3.46$ &   6340 & $-3.60$\\
 SDSS J$085211+033945$ & 08:52:12 & $+03$:39:45& 1  &    6343 &   $-3.24$ &   6365 & $-2.92$\\
 SDSS J$085232+112331$ & 08:52:33 & $+11$:23:31&4,8 &    6206 &   $-3.45$ &   6211 & $-3.55$\\ 
 SDSS J$090533-020843$ & 09:05:33 & $-02$:08:44&4,8 &    5974 &   $-3.44$ &   5971 & $-2.35$\\
 SDSS J$090733+024608$ & 09:07:33 & $+02$:46:08& 2  &    5934 &   $-3.44$ &   5916 & $-3.19$\\
 SDSS J$091753+523005$ & 09:17:53 & $+52$:30:05&10  &    5858 &   $-3.56$ &   5853 & $-3.49$\\
 SDSS J$091913+232738$ & 09:19:13 & $+23$:27:38&4,8 &    5490 &   $-3.25$ &   5486 & $-3.44$\\
 SDSS J$092912+023817$ & 09:29:12 & $+02$:38:17&5,6 &    5894 &   $-4.97$ &   5889 & $-1.36$\\
 SDSS J$102915+172928$ & 10:29:15 & $+17$:29:28& 3  &    5811 &   $-4.73$ &   5795 & $-4.29$\\
 SDSS J$103402+070116$ & 10:34:03 & $+07$:01:17& 5  &    6224 &   $-4.01$ &   6229 & $-3.69$\\
 SDSS J$105002+242109$ & 10:50:02 & $+24$:21:10&10  &    5682 &   $-3.90$ &   5676 & $-2.70$\\
 SDSS J$112031-124638$ & 11:20:32 & $-12$:46:39&4,8 &    6502 &   $-3.27$ &   6525 & $-1.70$\\
 SDSS J$112211-114809$ & 11:22:12 & $-11$:48:09&4,8 &    6157 &   $-3.00$ &   6159 & $-2.24$\\
 SDSS J$113528+010848$ & 11:35:28 & $+01$:08:49& 2  &    6132 &   $-3.03$ &   6210 & $-2.67$\\
 SDSS J$113723+255354$ & 11:37:23 & $+25$:53:54& 5  &    6310 &   $-2.70$ &   6186 & $-2.34$\\
 SDSS J$114424-004658$ & 11:44:25 & $-00$:46:59&4,8 &    6412 &   $-2.80$ &   6428 & $-2.21$\\
 SDSS J$120441+120111$ & 12:04:41 & $+12$:01:12&4,8 &    5839 &   $-3.37$ &   5834 & $-3.55$\\
 SDSS J$122935+262445$ & 12:29:36 & $+26$:24:46& 2  &    6452 &   $-3.29$ &   6496 & $-3.25$\\
 SDSS J$123055+000546$ & 12:30:55 & $+00$:05:47&4,8 &    6223 &   $-3.24$ &   6228 & $-3.22$\\
 SDSS J$123404+134411$ & 12:34:05 & $+13$:44:11&4,8 &    5659 &   $-3.59$ &   5653 & $-3.60$\\
 SDSS J$124304-081230$ & 12:43:04 & $-08$:12:31& 10 &    5488 &   $-3.48$ &   5484 & $-3.96$\\
 SDSS J$124502-073847$ & 12:45:03 & $-07$:38:47&4,8 &    6110 &   $-3.21$ &   6093 & $-2.91$\\
 SDSS J$130017+263238$ & 13:00:17 & $+26$:32:39& 5  &    6393 &   $-3.65$ &   6467 & $-3.11$\\
 SDSS J$131249+001315$ & 13:12:50 & $+00$:13:16&4,8 &    6400 &   $-2.38$ &   6416 & $-3.05$\\
 SDSS J$131456+113753$ & 13:14:57 & $+11$:37:53&4,8 &    6265 &   $-3.24$ &   6273 & $-3.39$\\
 SDSS J$132508+222424$ & 13:25:09 & $+22$:24:25&4,8 &    6292 &   $-2.60$ &   6301 & $-2.77$\\
 SDSS J$133718+074536$ & 13:37:19 & $+07$:45:36& 1  &    6377 &   $-3.48$ &   6398 & $-1.76$\\
 SDSS J$134922+140736$ & 13:49:23 & $+14$:07:37& 5  &    6112 &   $-3.60$ &   6112 & $-2.95$\\
 SDSS J$135331-032930$ & 13:53:31 & $-03$:29:30&4,8 &    6224 &   $-3.18$ &   6229 & $-3.07$\\
 SDSS J$141249+013206$ & 14:12:49 & $+01$:32:07&4,8 &    5799 &   $-2.94$ &   5792 & $-3.32$\\
 SDSS J$143632+091831$ & 14:36:32 & $+09$:18:31& 2  &    6340 &   $-3.40$ &   6445 & $-3.73$\\
 SDSS J$144640+124917$ & 14:46:41 & $+12$:49:17& 2  &    6189 &   $-3.16$ &   6222 & $-2.86$\\
 SDSS J$153747+281404$ & 15:37:48 & $+28$:14:05&4,8 &    6272 &   $-3.39$ &   6280 & $-3.55$\\
 SDSS J$154246+054426$ & 15:42:47 & $+05$:44:26& 2  &    6179 &   $-3.48$ &   6319 & $-3.16$\\
 SDSS J$154746+242953$ & 15:47:47 & $+24$:29:53&4,8 &    5903 &   $-3.16$ &   5898 & $-3.19$\\
 SDSS J$155751+190306$ & 15:57:52 & $+19$:03:06&4,8 &    6176 &   $-2.94$ &   6180 & $-2.59$\\
 SDSS J$200513-104503$ & 20:05:13 & $-10$:45:03&4,8 &    6289 &   $-3.41$ &   6298 & $-3.63$\\
 SDSS J$215023+031928$ & 21:50:24 & $+03$:19:28&4,8 &    6026 &   $-2.84$ &   6024 & $-3.40$\\
 SDSS J$215805+091417$ & 21:58:06 & $+09$:14:17&4,8 &    5942 &   $-3.41$ &   5938 & $-3.53$\\
 SDSS J$220121+010055$ & 22:01:22 & $+01$:00:56&4,8 &    6392 &   $-3.03$ &   6408 & $-0.47$\\
 SDSS J$220728+055658$ & 22:07:28 & $+05$:56:59&4,8 &    6096 &   $-3.26$ &   6096 & $-3.58$\\
 SDSS J$222130+000617$ & 22:21:30 & $+00$:06:17&4,8 &    5891 &   $-3.14$ &   5886 & $-2.93$\\
 SDSS J$223143-094834$ & 22:31:44 & $-09$:48:34& 2  &    6053 &   $-3.20$ &   6038 & $-3.01$\\
 SDSS J$230814-085526$ & 23:08:15 & $-08$:55:26& 2  &    6015 &   $-3.01$ &   6011 & $-2.32$\\
 SDSS J$231031+031847$ & 23:10:32 & $+03$:18:48&4,8 &    6229 &   $-2.91$ &   6235 & $-2.47$\\
 SDSS J$231755+004537$ & 23:17:56 & $+00$:45:38&4,8 &    5918 &   $-3.54$ &   5914 & $-3.41$\\
 SDSS J$233113-010933$ & 23:31:13 & $-01$:09:33& 2  &    6246 &   $-3.08$ &   6247 & $-3.39$\\
\hline
\end{longtable}
\tablebib
{(1) \citet{xgto}, (2) \citet{bonifacio12}, (3) \citet{leo},  (4) \citet{topos1}, 
(5) \citet{toposII}, (6) \citet{toposIII}, (7) \citet{toposIV}, (8) \citet{toposV}, (9) \citet{francois18}, and
(10) \citet{francois2020}.}

\begin{longtable}{lccrcccc}
\caption{Metallicities of stars in the PASTEL catalogue \citep{pastel}, compared to those derived by \mygi\ on the SDSS spectra.
\label{hrp}
}\\
\hline
\hline
Star  &  RA & DEC &    Ref. & \teff &      [Fe/H]  &  \teff &  [M/H]\\
      & \multicolumn{2}{c}{2000}              &     & K     &       dex    &   K    &    dex \\
      &     &     &         & Lit.  &       Lit.   & adop.  & \mygi \\
\hline
\endfirsthead
\caption{continued.}\\
\hline
\hline
Star  &  RA & DEC &    Ref. & \teff &      [Fe/H]  &  \teff &  [M/H]\\
      & \multicolumn{2}{c}{2000}             &     & K     &       dex    &   K    &    dex \\
      &     &     &         & Lit.  &       Lit.   & adop.  & \mygi \\
\hline
\endhead
\hline
\endfoot
 SDSS J000219.87+292851.7 & 00:02:20 & +29:28:52 & 1  & 6150   &$ -3.26 $&  6199 &$-2.84$  \\
     CS 22957-0036 & 00:03:31 & --04:44:23 &  2  & 5970         &$ -2.28 $&  6292 &$-2.07$   \\
 SDSS J002015.45-004058.1 & 00:20:15 & --00:40:58 & 1  & 6450   &$ -2.81 $&  6411 &$-2.70$    \\
 SDSS J002756.76-190929.8 & 00:27:57 & --19:09:30 & 1  & 6550   &$ -2.71 $&  6474 &$-2.88$    \\
 2MASS J00285742-1015305 & 00:28:57 & --10:15:31 & 1  & 6150    &$ -2.81 $&  6056 &$-2.98$    \\
 SDSS J004150.22-095327.7 & 00:41:50 & --09:53:28 & 1  & 6000   &$ -2.81 $&  5968 &$-2.71$      \\
 SDSS J010026.70+004915.6 & 01:00:27 & +00:49:16 & 1  & 5550   &$ -3.18 $&  5469 &$-3.08$    \\
 SDSS J011150.32+144207.8 & 01:11:50 & +14:42:08 & 1  & 6350   &$ -2.8  $&  6136 &$-2.70$    \\
 2MASS J02091203+2120281 & 02:09:12 & +21:20:28 & 1  & 6250    &$ -2.89 $&  6267 &$-2.85$    \\
 SDSS J025453.33+332840.9 & 02:54:53 & +33:28:41 & 1  & 6200   &$ -2.82 $&  6338 &$-2.84$     \\
 SDSS J030839.27+050534.8 & 03:08:39 & +05:05:35 & 1  & 5950   &$ -2.19 $&  5881 &$-1.94$      \\
 SDSS J035111.26+102643.1 & 03:51:11 & +10:26:43 & 1  & 5450   &$ -3.18 $&  5726 &$-2.45$      \\
 SDSS J072725.15+160949.4 & 07:27:25 & +16:09:50 & 1  & 6600   &$ -2.92 $&  6249 &$-2.99$       \\
 2MASS J07464132+2831426 & 07:46:41 & +28:31:43 & 1  & 6100    &$ -2.75 $&  6024 &$-2.65$       \\
 SDSS J074859.88+175832.6 & 07:49:00 & +17:58:33 & 1  & 6100   &$ -2.6  $&  6300 &$-2.74$  \\
 SDSS J080428.21+515303.1 & 08:04:28 & +51:53:03 & 1  & 5950   &$ -3.01 $&  5855 &$-2.84$  \\
 SDSS J080917.05+090748.5 & 08:09:17 & +09:07:49 & 1  & 6150   &$ -3.38 $&  6121 &$-3.03$  \\
 2MASS J08145867+3337130 & 08:14:59 & +33:37:13 & 1  & 6450    &$ -3.28 $&  6547 &$-3.23$    \\
 2MASS J08175492+2641038 & 08:17:55 & +26:41:04 & 1  & 6050    &$ -2.85 $&  6038 &$-2.98$   \\
 2MASS J08273626+1052008 & 08:27:36 & +10:52:01 & 1  & 6400    &$ -3.17 $&  6252 &$-3.39$     \\
 SDSS J084016.15+540526.4 & 08:40:16 & +54:05:26 & 1  & 6150   &$ -3.25 $&  6064 &$-2.86$  \\
 SDSS J084700.49+012113.6 & 08:47:00 & +01:21:14 & 1  & 6250   &$ -3.2  $&  6326 &$-3.09$   \\
 Cl* NGC 2682     ES    I-46 & 08:50:53 & +11:43:40 & 3 & 5703 &$ -0.07 $&  5534 &$-0.49$  \\
 Cl* NGC 2682    YBP    1514 & 08:51:01 & +11:53:11 & 4 & 5076 &$  0.09 $&  5520 &$-0.41$   \\
 SDSS J085136.67+101803.1 & 08:51:37 & +10:18:03 & 1  & 6500   &$ -2.89 $&  6290 &$-3.11$   \\
 SDSS J091243.72+021623.6 & 09:12:44 & +02:16:24 & 1  & 6150   &$ -2.68 $&  6097 &$-2.60$    \\
 SDSS J093247.29+024123.8 & 09:32:47 & +02:41:24 & 1  & 6200   &$ -3.14 $&  6202 &$-2.66$    \\
 SDSS J100427.70+344245.7 & 10:04:28 & +34:42:46 & 1  & 6100   &$ -3.09 $&  6091 &$-2.97$     \\
 2MASS J10330141+4001035 & 10:33:01 & +40:01:04 & 1  & 6600    &$ -3.06 $&  6391 &$-2.64$    \\
      SDSS J1036+1212 & 10:36:50 & +12:12:20 & 1  & 5850       &$ -3.47 $&  5833 &$-3.08$    \\
 SDSS J110610.48+034321.9 & 11:06:10 & +03:43:22 & 1  & 6300   &$ -2.88 $&  6203 &$-2.82$     \\
 2MASS J11082167+1747468 & 11:08:22 & +17:47:47 & 1  & 6050    &$ -3.17 $&  5971 &$-2.86$     \\
 SDSS J112051.73+302724.4 & 11:20:52 & +30:27:25 & 1  & 5750   &$ -3.14 $&  5752 &$-2.97$    \\
 SDSS J112813.56+384148.9 & 11:28:14 & +38:41:49 & 5 & 6449    &$ -3.53 $&  6616 &$-2.96$    \\
 SDSS J114723.53+151044.7 & 11:47:24 & +15:10:45 & 1  & 6500   &$ -2.96 $&  6528 &$-2.93$    \\
 SDSS J115906.18+542512.6 & 11:59:06 & +54:25:13 & 1  & 5700   &$ -3.26 $&  5724 &$-3.14$     \\
 SDSS J120441.38+120111.5 & 12:04:41 & +12:01:12 & 6 & 5467    &$ -4.34 $&  5834 &$-3.55$     \\
 SDSS J121307.22+445040.9 & 12:13:07 & +44:50:41 & 1  & 6650   &$ -3.2  $&  6411 &$-3.79$     \\
 SDSS J123055.25+000546.9 & 12:30:55 & +00:05:47 & 1  & 6150   &$ -3.34 $&  6228 &$-3.22$     \\
 2MASS J12330008+3407583 & 12:33:00 & +34:07:58 & 1  & 6300    &$ -2.87 $&  6175 &$-2.62$      \\
 SDSS J124502.68-073847.1 & 12:45:03 & --07:38:47 & 1  & 6100   &$ -3.17 $&  6093 &$-2.91$       \\
 2MASS J13001719+2632385 & 13:00:17 & +26:32:39 & 1  & 6450    &$ -3.53 $&  6467 &$-3.40$       \\
 SDSS J130402.25+323909.1 & 13:04:02 & +32:39:09 & 1  & 6050   &$ -2.9  $&  6075 &$-2.32$       \\
 SDSS J131201.47+245006.9 & 13:12:01 & +24:50:07 & 1  & 6250   &$ -2.72 $&  6193 &$-2.86$       \\
 SDSS J133453.44+002238.6 & 13:34:53 & +00:22:38 & 1  & 5650   &$ -3.03 $&  5636 &$-3.00$       \\
 SDSS J133841.16+120415.2 & 13:38:41 & +12:04:15 & 1  & 6300   &$ -2.86 $&  6284 &$-2.80$       \\
      SDSS J1349-0229 & 13:49:14 & -02:29:43 & 1  & 6200       &$ -3.24 $&  6138 &$-1.25$        \\
 SDSS J140035.31+075317.7 & 14:00:35 & +07:53:18 & 1  & 6250   &$ -2.98 $&  6269 &$-3.01$        \\
 SDSS J140813.88+623942.1 & 14:08:14 & +62:39:42 & 1  & 6300   &$ -2.97 $&  6267 &$-2.66$         \\
 2MASS J14100174+5350180 & 14:10:02 & +53:50:18 & 1  & 6100    &$ -3.42 $&  6109 &$-2.93$         \\
 SDSS J141207.32+560931.9 & 14:12:07 & +56:09:32 & 1  & 6600   &$ -3.19 $&  6413 &$-3.21$        \\
 SDSS J142441.88+561535.0 & 14:24:42 & +56:15:35 & 1  & 6350   &$ -2.97 $&  6234 &$-2.61$         \\
 SDSS J142518.09+113713.9 & 14:25:18 & +11:37:14 & 1  & 6300   &$ -3.08 $&  6253 &$-3.01$         \\
 SDSS J142541.33+574207.5 & 14:25:41 & +57:42:08 & 1  & 6450   &$ -3.29 $&  6343 &$-3.39$          \\
 SDSS J143451.02+103626.4 & 14:34:51 & +10:36:26 & 1  & 6400   &$ -3.21 $&  6319 &$-3.40$           \\
 SDSS J143654.45+030143.2 & 14:36:54 & +03:01:43 & 1  & 6150   &$ -3.6  $&  6180 &$-3.10$               \\
 SDSS J143708.91+523146.7 & 14:37:09 & +52:31:47 & 1  & 6300   &$ -2.9  $&  6227 &$-2.81$           \\
 SDSS J143759.06+583723.6 & 14:37:59 & +58:37:24 & 1  & 6450   &$ -3.02 $&  6384 &$-2.98$            \\
 SDSS J150217.16+311316.5 & 15:02:17 & +31:13:16 & 1  & 6350   &$ -2.86 $&  6181 &$-2.89$           \\
 SDSS J151534.44+450317.7 & 15:15:34 & +45:03:18 & 1  & 6300   &$ -3.34 $&  6227 &$-3.06$                     \\
 SDSS J151646.69+433331.6 & 15:16:47 & +43:33:32 & 1  & 6500   &$ -2.91 $&  6403 &$-2.98$    \\
 2MASS J15215863+3437292 & 15:21:59 & +34:37:29 & 1  & 6200    &$ -2.79 $&  6127 &$-2.97$       \\
 SDSS J152301.85+494210.6 & 15:23:02 & +49:42:11 & 1  & 5850   &$ -3.06 $&  5764 &$-2.83$     \\
 SDSS J152810.51+491526.8 & 15:28:11 & +49:15:27 & 1  & 6450   &$ -2.99 $&  6467 &$-2.84$     \\
 2MASS J15531082+2511404 & 15:53:11 & +25:11:40 & 1  & 5850    &$ -3.3  $&  5852 &$-2.88$      \\
 SDSS J160303.73+291709.5 & 16:03:04 & +29:17:10 & 1  & 6000   &$ -3.36 $&  6071 &$-2.94$             \\
 SDSS J162311.84+391319.6 & 16:23:12 & +39:13:20 & 1  & 6350   &$ -3.19 $&  6253 &$-2.78$           \\
 SDSS J162603.61+145844.3 & 16:26:04 & +14:58:44 & 1  & 6400   &$ -2.99 $&  6327 &$-2.18$          \\
 SDSS J163331.43+390742.6 & 16:33:31 & +39:07:43 & 1  & 6300   &$ -2.88 $&  6237 &$-3.02$           \\
 SDSS J164005.30+370907.8 & 16:40:05 & +37:09:08 & 1  & 6450   &$ -3.39 $&  6382 &$-3.24$          \\
 SDSS J164610.19+282422.2 & 16:46:10 & +28:24:22 & 1  & 6100   &$ -3.05 $&  6163 &$-2.28$           \\
 SDSS J165016.66+224213.9 & 16:50:17 & +22:42:14 & 1  & 6600   &$ -2.56 $&  6381 &$-3.03$           \\
 2MASS J16593474+3515545 & 16:59:35 & +35:15:54 & 1  & 6050    &$ -3.24 $&  6182 &$-2.74$            \\
 SDSS J172846.88+065701.9 & 17:28:47 & +06:57:02 & 1  & 6350   &$ -2.85 $&  6578 &$-3.31$               \\
 SDSS J183045.75+414126.8 & 18:30:46 & +41:41:27 & 1  & 6550   &$ -3.01 $&  6546 &$-3.77$                \\
 SDSS J183414.28+202335.5 & 18:34:14 & +20:23:35 & 1  & 6550   &$ -1.98 $&  6353 &$-2.03$                    \\
 2MASS J20051348-1045030 & 20:05:13 & --10:45:03 & 1  & 6600    &$ -3.34 $&  6298 &$-3.63$                      \\
 SDSS J205252.68+010939.3 & 20:52:53 & +01:09:39 & 1  & 6050   &$ -2.8  $&  6032 &$-2.68$               \\
 SDSS J211850.12-064055.8 & 21:18:50 & --06:40:56 & 1  & 6650   &$ -2.91 $&  6436 &$-2.73$                \\
 2MASS J21231081-0820393 & 21:23:11 & --08:20:39 & 1  & 6350    &$ -2.88 $&  6172 &$-3.01$                \\
 SDSS J212841.25-075629.3 & 21:28:41 & --07:56:29 & 1  & 6150   &$ -2.76 $&  6067 &$-2.85$                 \\
 SDSS J220743.35+201752.3 & 22:07:43 & +20:17:52 & 1  & 6200   &$ -2.42 $&  6433 &$-1.86$               \\
 SDSS J220845.57+061341.3 & 22:08:46 & +06:13:41 & 1  & 6450   &$ -2.85 $&  6300 &$-2.68$             \\
 V* FV Aqr & 22:12:03 & --08:45:46 &  2  & 6200                 &$ -2.58 $&  6568 &$-2.21$       \\
     CS 22886-0012 & 22:13:26 & --08:43:43 &  2  & 5650         &$ -2.88 $&  6056 &$-2.55$           \\
 SDSS J230026.35+055956.2 & 23:00:26 & +05:59:56 & 1  & 6450   &$ -2.94 $&  6343 &$-2.93$                 \\
 SDSS J233403.22+153829.3 & 23:34:03 & +15:38:29 & 1  & 6550   &$ -2.91 $&  6196 &$-2.81$           \\
 SDSS J234939.71+383217.8 & 23:49:40 & +38:32:18 & 1  & 6250   &$ -3.2  $&  6284 &$-2.67$                                  \\
\hline
\end{longtable}
\tablebib
{(1) \citet{aoki13}, (2) \citet{roederer14},(3) \citet{santos09}, 
(4) \citet{sousa18},  
(5) \citet{yong13}, 
(6) \citet{placco15}}

\section{Metallicity estimates from the $p$ index\label{appc}}

The procedure is as follows:

\begin{enumerate}
\item For each star, the effective temperature is estimated from the $(g-z)_0$ colour
using formula given in footnote\,\ref{tformula}, and the $p$ index is computed.
\item  The values of theoretical bolometric corrections in the $G$ band and $p$
indices are read for a set of eight effective temperatures and
six surface gravities, and nine metallicities are read.
\item For each star,   the parallax (obtained as $1/r_{est}$, where
$r_{est}$ is the geometrical distance estimate from \citealt{BJ}),
the effective temperature, and the $G0$ magnitude is read.
\item For an initial estimate of log g = 4.2, a first estimate
of metallicity is derived from the $p$ index, as well as
a first estimate of the bolometric correction,
by a spline interpolation in the theoretical table.
\item \label{iterg} A new gravity estimate is derived using the current 
bolometric correction estimate. 
\item \label{newmet} A new estimate of metallicity and bolometric correction is derived
with the current surface gravity.
\item Steps \ref{iterg} and \ref{newmet} are iterated until
the log g difference from the previous step is less than $0.05$\,dex.
\end{enumerate}

\end{document}